\documentclass[%
 aps,
 prx,
 reprint,
 amsmath,
 amssymb,
 amsthm,
 showkeys,
 superscriptaddress,
 eprint,
 nofootinbib,
 bibnotes,
 twoside
]{revtex4-2}

\usepackage[utf8]{inputenc}
\usepackage{graphicx}
\usepackage{xcolor}
\usepackage{enumerate}
\usepackage{amsfonts,amsmath,amssymb,amsthm}
\usepackage{dsfont}
\usepackage{mathrsfs}

\usepackage{hyperref}
\hypersetup{
	colorlinks = true,
	linkcolor = blue,
	citecolor = blue,
	urlcolor = blue,
	filecolor = black,
	linktoc = all,
}
\usepackage[numbered,openlevel=1,open]{bookmark}

\newcommand{\tr}{\operatorname{Tr}}
\newcommand{\one}{\mathds{1}}

\newcommand{\rr}{ {\hspace{1pt} \| \hspace{1pt}} }
\newcommand{\lam}{\lambda}

\newcommand{\MM}{\mathcal{M}}

\newcommand{\NN}{\mathcal{N}}
\newcommand{\AAA}{\mathcal{A}}
\newcommand{\BBB}{\mathcal{B}}
\newcommand{\PP}{\mathcal{P}}
\newcommand{\RR}{\mathcal{R}}
\newcommand{\FF}{\mathcal{F}}
\newcommand{\EE}{\mathcal{E}}

\newcommand{\chilostar}{\text{LO*}}
\newcommand{\chilo}{\text{LO}}
\newcommand{\chisep}{\text{SEP}}
\newcommand{\chippt}{\text{PPT}}
\newcommand{\chirct}{\text{RCT}}

\newcommand{\sep}{{\textsc{sep}}}
\newcommand{\locc}{{\textsc{locc}}}
\newcommand{\lo}{{\textsc{lo}}}
\newcommand{\lostar}{{\textsc{lo}^*}}
\newcommand{\ppt}{{\textsc{ppt}}}
\newcommand{\rct}{{\textsc{rct}}}

\newcommand{\req}{{\textsc{req}}}
\newcommand{\ree}{{\textsc{ree}}}

\newcommand{\cb}{{\textsc{cb}}}

\newcommand{\ent}{{\rm ent}}

\newcommand{\ie}{{\textit{i.e.}}}
\newcommand{\eg}{{\textit{e.g.}}}

\newcommand{\ket}[1]{|{#1}\rangle}
\newcommand{\bra}[1]{\langle{#1}|}
\newcommand{\ketbra}[1]{{\ket{#1} \! \bra{#1}}}

\newtheorem{thm}{Theorem}
\newtheorem{defn}[thm]{Definition}
\newtheorem{cor}[thm]{Corollary}
\newtheorem{rem}[thm]{Remark}
\newtheorem{lem}[thm]{Lemma}
\newtheorem{prop}[thm]{Proposition}

\begin{document}


\title{Observational entropy of quantum correlations and entanglement}


\author{Leonardo Rossetti}
\email{leonardo.rossetti@unicam.it}
\affiliation{School of Science and Technology, University of Camerino, Via Madonna delle Carceri 9, I-62032 Camerino, Italy}
\affiliation{Istituto Nazionale di Fisica Nucleare, Sezione di Perugia, via A. Pascoli, I-06123 Perugia, Italy}

\author{Stefano Mancini}
\email{stefano.mancini@unicam.it}
\affiliation{School of Science and Technology, University of Camerino, Via Madonna delle Carceri 9, I-62032 Camerino, Italy}
\affiliation{Istituto Nazionale di Fisica Nucleare, Sezione di Perugia, via A. Pascoli, I-06123 Perugia, Italy}

\author{Andreas Winter}
\email{andreas.winter@uni-koeln.de}
\affiliation{Department Mathematik/Informatik---Abteilung Informatik, Universit\"at zu K\"oln, Albertus-Magnus-Platz, D-50923 K\"oln, Germany}
\affiliation{F\'{\i}sica Te\`{o}rica: Informaci\'{o} i Fen\`{o}mens Qu\`{a}ntics, Departament de F\'{\i}sica, Universitat Aut\`{o}noma de Barcelona, ES-08193 Bellaterra, Spain}
\affiliation{ICREA, Passeig Llu\'{\i}s Companys, 23, ES-08010 Barcelona, Spain}
\affiliation{Institute for Advanced Study, Technische Universit\"at M\"unchen, Lichtenbergstra{\ss}e 2a, D-85748 Garching, Germany}

\author{Joseph Schindler}
\email{josephc.schindler@uab.cat}
\affiliation{F\'{\i}sica Te\`{o}rica: Informaci\'{o} i Fen\`{o}mens Qu\`{a}ntics, Departament de F\'{\i}sica, Universitat Aut\`{o}noma de Barcelona, ES-08193 Bellaterra, Spain}

\date{10 October 2025}

\begin{abstract}
The use of coarse graining to connect physical and information theoretic entropies has recently been given a precise formulation in terms of ``observational entropy'', describing entropy for observers with respect to a measurement. Here we consider observers with various locality restrictions, including local measurements (LO), measurements based on local operations with classical communication (LOCC), and separable measurements (SEP), with the idea that the ``entropy gap'' between the minimum locally measured observational entropy and the von Neumann entropy quantifies quantum correlations in a given state. 
After introducing entropy gaps for general classes of measurements and deriving their general properties, we specialize to LO, LOCC, SEP and other measurement classes related to the locality of subsystems. For those, we show that the entropy gap can be related to well-known measures of entanglement or non-classicality of the state (even though we point out that they are not entanglement monotones themselves). In particular, for bipartite pure states, all of the ``local'' entropy gaps reproduce the entanglement entropy, and for general multipartite states they are lower-bounded by the relative entropy of entanglement. The entropy gaps of the different measurement classes are ordered, and we show that in general (mixed and multipartite states) they are all different. 
\end{abstract}


\maketitle

\section{Introduction}
\label{sec:introduction}

Observational entropy (OE) is a measure of the uncertainty about a system's state that incorporates the specific measurement an observer performs to acquire information about the state~\cite{safranek2019a,safranek2019b,safranek2020classical,safranek2021brief}. Originally proposed by von Neumann as a notion of entropy more applicable to thermodynamic contexts than the previously introduced and more widely used von Neumann entropy~\cite{vonNeumann1929proof,tolman1938book,vonNeumann1955mathematical}, OE has since emerged as a valuable framework for studying classical and quantum out-of-equilibrium thermodynamics~\cite{strasberg2020first,strasberg2022book,riera2020finite,safranek2020quantifying,stokes2022nonconjugate,modak2022anderson,sreeram2022chaos,safranek2022work,sinha2023generalized,schindler2023continuity,bai2023observational,strasberg2023classicality,Strasberg_2024,nagasawa2024generic,safranek2023expectation,chakraborty2025sample,sreeram2025dichotomy,bhattacharjee2025work,schindler2025unification}. 

The intuition behind observational entropy lies in interpreting measurements as a coarse-graining of phase space or Hilbert space into ``macrostates'' corresponding to possible outcomes~\cite{boltzmann1909further,gibbs1902book,wehrl1979relation,goldstein2020gibbs}.  It is based on the idea that, even after obtaining a measurement outcome, residual uncertainty persists about the precise ``point'' (microstate) within the measured macrostate~\cite{safranek2019a,safranek2019b,safranek2020classical,safranek2021brief}. In principle, OE will depend on the chosen coarse-graining. However, when considering sets of coarse-grainings that share certain characteristics, it is possible to derive general properties of observational entropy that hold for the entire set under consideration~\cite{schindler2020correlation,piani2009relative,buscemi2022observational,bonfill2023entropic}.
Concretely, in a previous work~\cite{schindler2020correlation} it was shown that a general multipartite quantum state will exhibit a gap between the minimum locally-measured and globally-measured OE, for observers who are only capable of doing local projective measurements on their local subsystems. This gap was shown to be of interest as a bound on thermodynamic entropies and also as a measure of quantum correlations~\cite{schindler2020correlation,zhou2022relations,zhou2023dynamical}, and was found to coincide with other such measures~\cite{modi2010unified,groisman2007quantumness,piani2011nonclassical,horodecki2005local,oppenheim2002thermodynamical,horodecki2003local}. This entropy optimization problem is also closely related to studies of measured divergences under restricted measurements that have been of recent interest in quantum information theory~\cite{piani2009relative,mosonyi2023test,berta2024entanglement,rippchen2024locally}, of which the OE is a more easily analyzed special case. This state of affairs begs the question of how the observational entropy gaps generalize to other classes of restricted measurements. In particular, do the gaps associated with separable measurements quantify entanglement, as the gaps associated with local projective measurements quantify discord-like correlations? And what is the relative power of different local measurement classes (\eg~the utility of classical communication or non-projective measurements~\cite{calsamiglia2010local,matthews2009distinguishability}) in the context of measured information extraction?

In the present paper we thoroughly study the entropy gaps for different classes of measurements relating to locality restrictions, and for each class, we analyze the potential of the corresponding entropy gap for quantifying entanglement and quantum correlations. We find that using different classes provides us with diverse tools for measuring quantum correlations, each capable of capturing different aspects of these correlations. Moreover, depending on the class considered, we are able to establish connections with various known measures of quantum correlation and entanglement.

We start by providing a definition of entropy gap as generally as possible: given a set $\chi$ of measurements, we define $S_\chi$ as the minimal OE over $\chi$, and the corresponding entropy gap $E_\chi$ as the difference between $S_\chi$ and $S$, the von Neumann entropy. We establish several general properties of these quantities. 
These properties are then exploited in the core of our work, where we study the entropy gaps for classes of measurements subject to specific locality constraints. In addition to the class of local projective measurements (LO*), we consider the class of general local measurements (LO), the class of measurements based on local operations and classical communication (LOCC), and the class of separable measurements (SEP). Sometimes it is convenient to use further relaxations of SEP, which we will mention in passing where appropriate: measurements whose operators have positive partial transpose (PPT) and those which satisfy the reduction criterion (RCT), cf.~\cite{horodecki2009entanglement}.
As mentioned already, each entropy gap associated with one of these classes represents a distinct measure of quantum correlation.

Before launching into the technical part of the paper, we provide an overview of our main results.
\begin{itemize}

    \item We show that for any separable measurement $M$ (or even any $M \in \chirct$), the observational entropy $S_M(\rho_{AB})$ of a bipartite state is $\geq$ the marginal von~Neumann entropy $S(\rho_A)$ (or likewise $B$). This provides our fundamental bound on information extraction by locality-restricted measurements. This also  marks a crucial difference between observational and von Neumann entropy, making OE more akin to a classical entropy. As a corollary, for any set $\chi$ where $\chilostar \subset \chi \subset \chirct$, the entropy~gap $E_\chi(\ket{\psi_{AB}}) = S(\rho_A)=S(\rho_B)$ is equal to the entanglement entropy on all bipartite pure states.

    \item We prove that $E_\sep$ is $\geq$ the relative entropy of entanglement, making use of Petz recovery maps and the OE recovery bound~\cite{buscemi2022observational}. We also confirm that the gap $E_\lostar$ is equal to the relative entropy of quantumness (discord)~\cite{schindler2020correlation,modi2010unified}. The local gaps are therefore bounded below by the amount of entanglement and from above by the amount of discord.
    
    \item We precisely characterize the set of states $\rho$ for which \mbox{$E_\chi(\rho)=0$}. We show that $E_\lo(\rho) = 0$ if and only if $E_\lostar(\rho)=0$. In particular we find the LO* and LO gaps are zero on the set of classically correlated states. The SEP gap is found to be zero on a set of states that are called ``eigenseparable'', and we discuss the characteristics of this set (one interesting feature is that, to be eigenseparable, a state not only has to be separable, but also have a separable 0-eigenspace projector). We exhibit separable states which are not eigenseparable, demonstrating that some states which can be prepared locally cannot be optimally measured by local measurements.

    \item We prove that $S_\lostar \neq S_\lo \neq S_\locc \neq S_\sep$ in general, by virtue of specific counterexamples to equality. Of course there are certain $\rho$ for which the equalities hold, but in general the measures are separated. The separation examples help illustrate what kind of information tasks require the increasingly powerful measurement strategies like POVMs or classical communication, compared to traditional projective measurements. In particular we use a classical-quantum trine state to prove that local POVMs (LO) are strictly more powerful than local projective measurements (LO*), which is especially interesting because when they equal \textit{zero} the two measures are equivalent. It is then easy to separate LO from the much more powerful LOCC. An example based on states from \cite{bennet1999nonlocality} separates LOCC from SEP, an example based on unextendible product bases separates SEP from PPT, and a Werner state separates PPT from RCT.

    \item We study monotonicity of the local entropies under quantum channels. We find $S_\sep(\Phi(\rho)) \geq S_\sep(\rho)$ for all unital separable channels $\Phi$. Similarly, $S_\lo(\Phi(\rho)) \geq S_\lo(\rho)$ for all unital local channels. This suggests a natural setting for these entropies might be a joint resource theory of purity (unital) and entanglement (separable). However, the full resource theoretic picture remains unclear. In particular we are able to show by counterexamples that the gaps $E_\sep$ and $E_\lo$ are \textit{not} monotones of these same classes. Moreover, we exhibit separable Werner states such that all the local gaps are strictly $> 0$. This shows that none of the local gaps is an entanglement monotone. Rather they quantify more general quantum correlations, with the relation to entanglement measures being (as noted earlier) an inequality.

    \item We discuss the problem of identifying optimal measurements for entropy minimization on state $\rho$, within each class. We obtain some partial strategies such as the optimal LO/LO* strategy for CQ (classical-quantum) states, which we prove requires measuring the C side in its classical basis. For such states we relate the entropy optimization problem to the accessible information and Holevo's bound. Meanwhile, we reduce the LO* problem to searching for a basis, and the SEP problem to its rank-1 refinement. We discuss a chain rule for one-way LOCC strategies and provide an upper bound on the usefulness of additional LOCC rounds. We also exhibit different entangled/separable/discordant states and how they are quantified by the measures.

    \item We utilize the gaps to study multipartite aspects of quantum correlations. For this it is relevant that the gaps are on the same scale regardless of how a multiparty system is partitioned, and can be meaningfully compared across partition types, with a monotonicity under refining the partition. We compare 4-party GHZ, W, and double-Bell states, evaluating the LO* gap in all possible partitions. We find that among these, W states have the more robust and more genuinely multipartite correlations, as seen in Fig.~\ref{fig:multipartite-genuine-robust}.

    \item We evaluate the gaps in a number of examples: We analytically calculate the local gaps on all Werner states, shown in Fig.~\ref{fig:werner-plots}. We give a formula for LO/LO* gaps for CQ states and use it to prove that $E_\lostar(\rho)=0.5$ bits for the state $\rho = (\ketbra{00} + \ketbra{1+})/2$, which was previously known numerically, and show that $E_\locc = 0$ for all CQ states. We compare pure entangled states to their dephased CQ counterparts, and show that, of the two, $E_\chi$ gaps are larger for the pure state. We also evaluate the multipartite W, GHZ, and Bell examples already mentioned above. Finally we study the local gaps in W states in detail. For LO* it was found numerically that the computational basis measurement is optimal, giving $E_\lostar = \log n$ (with $n$ the number of parties). Numerical study of the LO case revealed the same bound. However, even for the LO* case we did not obtain an analytical proof of the value. We describe an LOCC measurement that does do strictly better than $\log n$. And for PPT we are able to prove that $E_\ppt = \log \frac94$, lower-bounding the SEP and LOCC values.

\end{itemize}

In addition to the main results, we also provide some useful supporting analysis in the setting of general $\chi$, some parts of which are also new.
\begin{itemize}

    \item We discuss the problem of $E_\chi = 0$ in the general case and prove strong equivalent conditions for a measurement to be optimal. We also study elementary properties like convexity and bounds. 

    \item We establish monotonicity of OE under quantum post-processing (QPP, which makes measurements finer using conditional sequential measurements), and classical post-processing (CPP, which makes measurements coarser by relabeling and rebinning outcomes). We provide a direct chain rule for the QPP monotonicity. We define the ``post-processing completion'' of set $\chi$, which are the measurements practically achievable by creative use of a more basic primitive set. These methods are later used to define LOCC and obtain the LOCC chain rule.

    \item We prove a general monotonicity result that $S_\chi(\Phi(\rho)) \geq S_\chi(\rho)$ for any $\chi$-preserving unital channel $\Phi$ (as defined in the text). We review the OE recovery inequality, and use it to show $E_\chi$ is $\geq$ a measure we call the relative entropy of \mbox{$\chi$-ness}. This is later specialized to entanglement and discord cases.
\end{itemize}

The paper is structured as follows. Sec.~\ref{sec:oe-and-quantum-measurements} recalls the definition of observational entropy in its most general form, along with some of its most fundamental information-theoretic and statistical interpretations. In Sec.~\ref{sec:chi-entropy-gap} we define $S_\chi$ and $E_\chi$, whose general properties, for $\chi$ a given set of measurements, are explored in Sec.~\ref{sec:aspectsofechi}. Sec.~\ref{sec:multipartite} provides basic definitions and properties of measurements and observational entropy in multipartite settings. In Sec.~\ref{sec:local-measurement-classes}, we formally define the local measurement classes on which we will primarily focus, and prove the separable measurement theorem. Also there, in Subsec.~\ref{sec:entanglement-discord-and-re-measures}, we establish a connection between $E_\chi$ and other known measures of entanglement and quantum correlations, such as quantum discord and relative entropy of entanglement. In the subsequent Secs.~\ref{sec:lovslostar}, \ref{sec:locc} and \ref{sec:SEP} we analyze the LO and LO*, LOCC, and SEP classes, respectively. Sec.~\ref{sec:examples} presents explicit calculations of entropy gaps for specific cases, such as Werner states, classical-quantum states, and W states, providing among other things separations between the entropy gaps of all the classes investigated. We also discuss the genuineness of multipartite correlations, which is examined by analyzing the behavior of $E_{\lostar}$ in some specific examples. 
Finally, we conclude in Sec.~\ref{sec:conclusions}.

\section{Observational entropy and quantum measurements}
\label{sec:oe-and-quantum-measurements}

Consider a state (density operator) $\rho$ on a $d$-dimensional Hilbert space, \ie~$\rho = \rho^\dagger\geq 0$ and \mbox{$\tr\rho=1$}. Throughout the paper we shall deal with finite-dimensional systems, unless it is otherwise specified. A quantum measurement can be defined as a quantum instrument $\MM = (\MM_i)_{i}$, which is a collection of quantum operations $\MM_i$ (completely positive maps~\cite{hayashi2016quantum}) that satisfy the trace-preserving condition $\tr \sum_i \MM_i (\rho) = \tr \rho$. This definition of measurements encodes both the measurement outcome probabilities given by $p_i = \tr \MM_i(\rho)$, and the post-measurement states $\rho_i = \frac{1}{p_i}\MM_i(\rho)$ that result after the measurement. For example, the instrument $\MM_i(\bullet) = \ketbra{i} \bullet \ketbra{i}$, resulting in $p_i = \bra{i} \rho \ket{i}$ and $\rho_i = \ketbra{i}$, represents a measurement in the basis $\{\ket{i}\}_i$. The instrument definition allows the usual measurements of Hermitian observables to be generalized to include any possible measurement, including noisy measurements, measurements that make use of an ancilla system, or measurements where quantum or classical outcome data is stored in external registers. Moreover, quantum instruments also allow the description of sequential measurement processes, where post-measurement states are retained to be measured again later, such as occurs during LOCC protocols.

While quantum instruments provide a fully general framework for describing measurements, a simplified approach suffices when only outcome probabilities, rather than post-measurement states, are of interest. In such a context, each $\MM_i$ can be associated with a unique positive-semidefinite Hermitian operator $M_i$ such that the outcome probabilities are given by $p_i = \tr \MM_i(\rho) = \tr M_i \rho$; namely $M_i = \MM_i^\dagger(\one)$, where $\MM_i^\dagger$ is the adjoint superoperator of $\MM_i$~\cite{hayashi2016quantum}, and these $M_i$ form a POVM. Therefore, for describing outcome probabilities, general measurements can be represented by POVMs (positive-operator valued measures)~\cite{hayashi2016quantum}, defined as
\begin{equation}
    M=(M_i)_{i}, \qquad M_i \geq 0, \qquad \sum_i M_i = \one,
\end{equation}
\ie~collections of positive-semidefinite Hermitian operators summing to the identity. A POVM determines an associated instrument by taking as quantum operations $\MM_i(\bullet) = M_i^{1/2} \bullet M_i^{1/2}$ (this is known as L\"uders rule, generalizing the projection postulate). However, this determination is not unique---there are always many possible instruments sharing the same POVM.

Observational entropy can be defined with either instruments or POVMs defining the measurement. The instrument definition is somewhat redundant, as the entropy value depends only on the corresponding POVM. Nonetheless, it is useful to establish both notations.

\begin{defn}[Observational entropy (OE)]
Let $\rho$ be a state. The observational entropy of $\rho$ coarse-grained by a quantum instrument $\MM = (\MM_i)_i$ is defined as
\begin{equation}
    \label{eqn:oe-definition-instrument}
    S_\MM(\rho) \equiv -\sum_i p_i \log \frac{p_i}{V_i}.
\end{equation}
where $p_i = \tr \MM_i(\rho)$ and $V_i = \tr \MM_i(\one) = \tr \MM_i^\dagger(\one)$. Equivalently, the observational entropy of $\rho$ coarse-grained by the associated POVM $M = (M_i)_i$ is defined as
\begin{equation}
    \label{eqn:oe-definition-POVM}
    S_M(\rho) \equiv -\sum_i p_i \log \frac{p_i}{V_i},
\end{equation}
where $p_i = \tr \rho M_i$ and $V_i = \tr M_i$. In either case, $p_i$ is the probability of observing $\rho$ in the $i$-th ``macrostate'' (\ie, obtaining the measurement outcome $i$), and $V_i$ represents the corresponding macrostate ``volume'', generalizing the concept of classical phase space volumes.
\end{defn}

An important class of measurements is the set of projective measurements $\Pi = (\Pi_i)_{i}$, which are POVMs that obey the additional condition
\begin{equation}
    \Pi_i \Pi_j = \delta_{ij} \, \Pi_i,
\end{equation}
so that $(\Pi_i)_i$ represents a complete set of orthogonal projectors. A projective measurement is equivalent to a decomposition of the Hilbert space into linear subspaces, and measurements of any Hermitian observable fall into this class. Non-projective measurements are more general, allowing the possibility of coupling a system to an ancilla and measuring observables in the joint system. One question we will explore is whether non-projective local measurements can be more powerful than projective ones in extracting information from a system. 

The observational entropy has several equivalent characterizations that help shed light on its significance. One interpretation is via
\begin{equation}
  S_M(\rho)  = H(p) + \sum_i p_i \log V_i,
\end{equation}
which expresses OE as the sum of two terms: the uncertainty in the measurement outcomes, quantified by Shannon entropy $H(p) = -\sum_i p_i \log p_i$ of the induced outcome distribution, and the average uncertainty that remains after the measurement outcome is known, quantified by Boltzmann entropies $\log V_i$. By incorporating both these types of uncertainty, observational entropy naturally captures several different types of entropy increase, leading to its utility in describing second laws in non-equilibrium statistical mechanics~\cite{vonNeumann1929proof,wehrl1978general,safranek2021brief}. 

Another interpretation comes from the information-theoretic perspective. Defining the \textit{measured relative entropy}
\begin{equation}
    D_M(\rho \rr \sigma)\equiv D(p^\rho \rr p^\sigma)
\end{equation}
as the classical relative entropy between $p^\rho_i = \tr M_i \rho$ and $p^\sigma_i = \tr M_i \sigma$ (the probability distributions induced by measurement $M$ acting on $\rho$ and $\sigma$, respectively), the OE is given by
\begin{equation}
\label{eqn:oe-mre}
    S_M(\rho) = \log d - D_M(\rho \rr \one/d),
\end{equation}
with $d$ the dimension of the Hilbert space where $\rho$ lives. This expresses OE in terms of the information gained by performing $M$ on the system, given the prior assumption that all microstates were in principle equally likely. 

The informational form can be extended by introducing the quantum channel
\begin{equation}
    \Phi_M(\bullet) = \sum_i \tr(M_i \bullet) \ketbra{i},
\end{equation}
with $\{\ket{i}\}_i$ being a set of orthonormal vectors. The measured relative entropy can then be equivalently expressed as $D_M(\rho \rr \sigma) = D(\Phi_M(\rho) \rr \Phi_M(\sigma))$, so that the OE has both classical and quantum informational expressions.

\section{Entropy gaps of restricted measurement}
\label{sec:chi-entropy-gap}
For an arbitrary state $\rho$, the minimum value of observational entropy over all possible measurements is~\cite{vonNeumann1929proof,vonNeumann1955mathematical,safranek2021brief}
\begin{equation}
\label{eqn:von-neumann-min}
    \min_{M \text{ POVM}} S_M(\rho) = S(\rho),
\end{equation}
where $S(\rho) = -\tr \rho \log \rho$ is the von Neumann entropy. This shows that, regardless of the measurement performed, the ultimate knowledge an observer can have about a quantum state is determined by its inherent uncertainty, $S(\rho)$.

However, actual observers may not have the ability to perform every measurement that is theoretically possible, but rather will be limited to some available set $\chi$ of POVMs that they can in reality implement. Among the measurements $M \in \chi$, those that yield lower entropy are of greater utility to the observer: if one can identify a measurement $M$ revealing low entropy, they can extract resources from the system by manipulating that degree of freedom. 

In this sense, the minimal entropy over $\chi$ determines how entropic the state effectively is to such an observer. The gap in entropy, above the global minimum~\eqref{eqn:von-neumann-min}, is the amount of extra entropy such an observer witnesses due to the limitations imposed by the measurement set available to them. This motivates the general definition of entropy gaps.

\begin{defn}[Entropy gaps]
\label{entropygaps}
    Let $\chi$ be a set of POVMs and $\rho$ a state. The minimum observational entropy of $\rho$ over this set is
    \begin{equation}
    \label{eqn:SX}
        S_\chi(\rho) \equiv \inf_{M \in \chi} S_M(\rho),
    \end{equation}
    and the corresponding \textit{entropy gap} is defined by
    \begin{equation}
    \label{eqn:EX}
        E_\chi(\rho) \equiv S_\chi(\rho) - S(\rho).
    \end{equation}
    The latter is the excess minimal entropy above $S(\rho)$, which would be the minimum if all $M$ were possible.
\end{defn}

A fundamental result obtained in previous work~\cite{schindler2020correlation} (see also~\cite{piani2009relative,rippchen2024locally}) demonstrates that, when two parties are limited to local projective measurements on their respective subsystems, for bipartite pure states, the entropy gap corresponds to the entanglement entropy, a well-known measure of pure state entanglement. For mixed and multipartite states, the same gap was shown to generalize the entanglement entropy to a total measure of quantum correlations.

In this work, we extend the investigation to other local measurement classes of operational interest, such as LOCC (local operations and classical communication), and separable measurements. In contrast to the local projective measurements studied previously, which ultimately connect entropy gaps to measures of discord-type correlations~\cite{modi2010unified}, these larger classes are those associated with quantum entanglement~\cite{horodecki2009entanglement}.

\section{General aspects of entropy gaps}
\label{sec:aspectsofechi}
A motivation for focusing on local measurement classes stems from the fact that locality is among the most inevitable limitations any experimental setting must face. However, locality is not the only reason that available measurements may be limited, and any set of experimentally dictated laboratory limitations could be of equal interest. To include any particular case, and also to clarify the mathematical arguments used, we first develop some fundamental properties of the entropy gaps for a general set $\chi$ of measurements.

\subsubsection{Bounds and optimal measurements}

One elementary property is the range of values the gap may take, which obeys the same bounds as the value of an entropy of the system.
\begin{prop}[General bounds]
\label{prop:genbounds}
    Let $\chi$ be a set of measurements, and $\rho$ be a state. The gap $E_{\chi}(\rho)$ can take the range of values
        \begin{equation}
        \label{eqn:general-bounds}
            0\leq E_{\chi}(\rho) \leq \log d-S(\rho).
        \end{equation}
    These bounds can be saturated; for instance, if $\tau=\one/d$ is the maximally mixed state and $M_\one = (\one)$ the trivial measurement, then $E_{\{M_\one\}}(\rho) = \log d - S(\rho)$ for any $\rho$, and $E_\chi(\tau)=0$ for any $\chi$.
\end{prop}
\begin{proof}
The bounds follow from~\eqref{eqn:von-neumann-min} and from the elementary bounds $ 0 \leq S_M(\rho) \leq \log d$ on OE~\cite{safranek2020quantifying}. By direct calculation using \eqref{eqn:oe-mre}, $S_M(\tau)=S(\tau)=\log d$, and $S_{M_\one}(\rho)=\log d$, ensuring saturation.
\end{proof}

Determining the best measurement from a set $\chi$, for the purpose of minimizing $S_M(\rho)$ on some state, is in general a difficult problem. Typically neither the form of such a measurement, nor the value $E_\chi(\rho)$ of the gap, can be deduced from the state in a simple analytical form. 

A simpler question, however, is whether or not $\chi$ contains any globally optimal measurement, meaning one that saturates the lower bound $S_M(\rho)=S(\rho)$. This is equivalent to the question of for which $\rho$ does $E_\chi(\rho)=0$. For this purpose it is useful to state in a strong form the conditions for some $M$ to be OE optimal.

\begin{lem}[Optimal measurement]
\label{lemma:optimal-meas}
Let $M = (M_i)_{i \in I}$ be a POVM whose outcomes are indexed by the set $I$.
Let~$\rho$ be a state with spectral decomposition $\rho \! = \! \sum_k \! \lam_k \Pi_k$, with $\lam_k$ the distinct eigenvalues and $\Pi_k$ the projectors onto the corresponding eigenspaces. 
The following are equivalent:
\begin{enumerate}[~(a)]
    \item $S_M(\rho) = S(\rho)$; 
    \item For each $M_i$ there is a unique value $k(i)$ such that $M_i = \Pi_{k(i)} M_i \Pi_{k(i)}$, and for all other $k\neq k(i)$, $\Pi_k M_i = 0 = M_i \Pi_k$;
    \item The set of outcomes $I$ can be decomposed into disjoint subsets~$I(k)$ such that each $\Pi_k = \sum_{i \in I(k)} M_i$, with $M_i = \Pi_k M_i \Pi_k$ for all $i \in I(k)$;
    \item $M$ is a refinement of $M_\rho = (\Pi_k)_k$ (in the standard post-processing sense, see Thm.~\ref{thm:cpp-qpp-monotonicity} below), and thus $S_M(\sigma) \leq S_{M_\rho}(\sigma)$ for all $\sigma$.
\end{enumerate}
In other words, if $M$ is optimal then each POVM element is completely supported on one of the eigenspaces of $\rho$, each eigenprojector is a sum over POVM elements, and $M$ is finer than the measurement of eigenprojectors of $\rho$.
\end{lem}
\begin{proof}
    The proof is based on the equality condition for Jensen's inequality on the function $F(x) = - x \log x$, applied to $F(p_i/V_i)$ for each fixed measurement outcome. The equality condition ensures each $M_i$ is orthogonal to all but one of the $\Pi_k$ eigenprojectors, denoted $\Pi_{k(i)}$, which is used to show that $\Pi_{k(i)} M_i \Pi_{k(i)} = M_i$. This allows the outcome set $I$ to be decomposed based on which elements $i$ overlap with each $k$ eigenspace.

    In particular, let $p_i = \tr M_i \rho$ and $V_i = \tr M_i$, and begin by rewriting $S_M(\rho) = \sum_i V_i \, F(p_i/V_i)$. Define $K(i) = \{ k \, | \, \tr M_i \Pi_k \neq 0\}$, and let $\alpha_{ik}=\tr(M_i \Pi_k)/V_i$. Observe that $p_i/V_i = \sum_{k\in K(i)} \alpha_{ik} \lam_k$, with $\alpha_{ik}\geq 0$ and normalized over the $k$ sum. Jensen's inequality therefore states that $F(p_i/V_i) \geq \sum_{k \in K(i)} \alpha_{ik} F(\lam_k)$, with equality if and only if $\lam_k = \lam_{k'}$ for all $k,k' \in K(i)$, because $F(x)$ is strictly concave. If the equality conditions hold then $S_M(\rho) = \sum_k \tr \Pi_k \, F(\lam_k) = S(\rho)$, and this is true only if they hold since otherwise Jensen's inequality would be strict.

    Thus $S_M(\rho)=S(\rho)$ if and only if $\lam_k = \lam_{k'}$ for all $k,k' \in K(i)$. Since the $\lam_k$ are all distinct, this means $K(i)$ contains only a single value, denoted $k(i)$, and thus $\tr M_i \Pi_k = 0$ for all $k \neq k(i)$. But with $\Pi_k^2 = \Pi_k$ and cyclicity of trace, this implies that for all $k \neq k(i)$ one has $A_{ik} \equiv M^{1/2}_i \Pi_k = 0$ , since then $\tr(A_{ik}^\dag A_{ik})=0$. Thus $ \one M_i \one = \sum_{kk'} \Pi_k M_i \Pi_{k'} = \Pi_{k(i)} M_i \Pi_{k(i)}$, which proves (a) implies (b).

    Now define $I(k) = \{i \in I \, | \, \Pi_k M_i \Pi_k \neq 0 \}$. From (b) it follows that $\Pi_k M_i \Pi_k =  M_i \, \delta_{k,k(i)}$. Therefore the sets $I(k)$ are disjoint, and one has $M_i = \Pi_k M_i \Pi_k$ for all $i \in I(k)$. From completeness of $M$ it then follows that $\Pi_k = \Pi_k \one \Pi_k = \sum_{i \in I} \Pi_k M_i \Pi_k  = \sum_{i \in I(k)} M_i$, giving the remaining property. Thus (b) implies (c). 

    Given (c), in terms of Thm.~\ref{thm:cpp-qpp-monotonicity}, $M_\rho = \Lambda M$ under the stochastic map $\Lambda_{k|i} = \delta_{k,k(i)}$, so (c) implies (d). Finally, one can directly evaluate that $S_{M_{\rho}}(\rho) = S(\rho)$. Given (d), since this is the minimum, one has $S(\rho) \geq S_M(\rho) \geq S(\rho)$. Therefore (d) implies (a), completing the proof.
\end{proof}

As a final elementary bound, we consider convexity properties of the $E_\chi$ gap. For any fixed $M$, the gap $E_M(\rho) = S_M(\rho)-S(\rho)$ is the difference of two concave functions of $\rho$, but the concavities go in opposite directions. The following lemma, which follows from Holevo's bound~\cite{wilde2011notes}, shows that for fixed $M$, this difference is itself actually convex. However, the infimum goes in the wrong direction to utilize this convexity, so the gap over a nontrivial set $\chi$ does not inherit the convexity.

\begin{lem}[Convexity]
\label{thm:convexity}
    Let $\rho = \sum_k \lam_k \rho_k$ be a convex combination of states $\rho_k$. For any fixed POVM $M$,
    \begin{equation}
        E_M(\rho) \leq \textstyle\sum_k \lam_k \, E_M(\rho_k).
    \end{equation}
    Thus, if $\chi$ is a set of measurements,
    \begin{equation}
        E_\chi(\rho) \leq \inf_{M \in \chi} \; \textstyle\sum_k \lam_k \, E_M(\rho_k).
    \end{equation}
\end{lem}
\begin{proof}
    For fixed $M$, the difference $S_M(\rho) - \sum_k \lam_k S_M(\rho_k)$ is equal to the equivalent expression for the Shannon part of the entropy, and is therefore subject to Holevo's bound~\cite{wilde2011notes}. The resulting inequality can be rearranged to the above form, as also proved in~\cite[(38)]{schindler2025unification}. The case of $\chi$ immediately follows from writing $E_\chi$ as an infimum and using the fixed $M$ case.
\end{proof}

\subsubsection{Classical and quantum postprocessing}
\label{sec:general:cpp-qpp}

Among the most basic properties of observational entropy is that coarser measurements give higher entropy, and finer measurements give lower entropy.

Coarseness and fineness of POVMs is generally defined with respect to the postprocessing partial order, which states that $N$ is coarser than $M$ (and $M$ is finer than~$N$) if there exists a classical channel $\Lambda$ such that $N = \Lambda M$, in the sense of ``classical postprocessing'' defined below. This coarse/fine ordering of POVMs is reviewed \eg~in~\cite{martens1990nonideal,buscemi2005clean,buscemi2015degradable,leppajarvi2021postprocessing,bonfill2023entropic} (for broader discussion of measurement processing and transformations see also \cite{buscemi2020complete,buscemi2023unifying}).

Given some initial set $\chi_0$ of available measurements, ``postprocessing'' operations can be used to construct other coarser and finer ones, using a combination of sequential measurement and data processing steps.
 
Classical postprocessing (CPP) describes the process by which outcomes may be relabelled and rebinned, possibly stochastically, during classical data processing~\cite{leppajarvi2021postprocessing,guff2021resource}. This can be defined for either instruments or POVMs.

\begin{defn}[CPP]
\label{def:cpp}
     The classical postprocessing of instrument $\MM$ by a stochastic map $\Lambda$ is the instrument $\Lambda \MM$ whose elements are given by
     \begin{equation}
         (\Lambda \MM)_j = \sum_i \Lambda_{j|i} \, \MM_i.
     \end{equation}
     For $M$ a POVM, the CPP $\Lambda M$ is defined analogously. A stochastic map $\Lambda$ is a matrix such that $\Lambda_{j|i} \geq 0$ and $\sum_j \Lambda_{j|i}=1$, mapping the original set of outcomes $\{i\}_i$ into a new set of outcome labels $\{j\}_j$.
\end{defn}

Quantum postprocessing (QPP) describes a sequential measurement scenario where subsequent measurements are chosen conditionally based on earlier outcomes~\cite{leppajarvi2021postprocessing}. 

\begin{defn}[QPP]
\label{def:qpp}
    Let $\MM = (\MM_i)_{i}$ be a quantum instrument. For each of its possible outcomes ``i'',  let $\NN^{(i)}$ be a quantum instrument. Denote the collection of these instruments by $\NN = (\NN^{(i)})_{i}$, and denote the elements of each individual instrument by $\NN^{(i)} = (\NN_{j|i})_{j}$. The quantum postprocessing of $\MM$ by $\NN$ is the instrument
     \begin{equation}
             \NN : \MM = (\NN_{j|i} \circ \MM_i)_{i,j}
     \end{equation}
     Note that $\NN : \MM$ is an instrument, even though $\NN$ (being a collection of instruments) is not. Physically this corresponds to first performing instrument $\MM$, then conditionally, for whichever outcome ``i'' is obtained, performing the instrument $\NN^{(i)}$.
 \end{defn}

Because it involves post-measurement output states, QPP must be defined in terms of instruments. At the level of POVMs, the POVM $N : M$ resulting from instrument $\NN : \MM$ is always finer than the POVM $M$, since there exists a CPP $\Lambda$ (namely $\Lambda_{i|i'j'} = \delta_{ii'}$) that undoes the QPP, so that $\Lambda (N :M) = M$ relates the POVMs. Thus the two types of postprocessing have opposite behavior, with CPP/QPP making $M$ coarser/finer respectively.

\smallskip

A central property of OE is its monotonicity under these postprocessings. These are often referred to as coarser/finer monotonicity in the CPP case, and sequential measurement monotonicity in the QPP case~\cite{buscemi2022observational}.

\begin{lem}[CPP/QPP monotonicity]
\label{thm:cpp-qpp-monotonicity}
Let $\Lambda \MM$ and \mbox{$\NN : \MM$} be instruments defined by CPP and QPP respectively. Then for any state $\rho$, one has $S_{\Lambda\MM}(\rho) \geq S_{\MM}(\rho)$ and $ S_{\MM}(\rho) \geq S_{\NN:\MM}(\rho)$. Similarly, $S_{\Lambda M}(\rho) \geq S_M(\rho)$ for any POVM.
\end{lem}

\begin{proof}
    The proof for CPP is an instance of the monotonicity of relative entropy under classical stochastic channels, deriving from the log sum inequality~\cite{cover2006book}. The proof for QPP follows from the fact that $N : M$ is finer than $M$ in the CPP sense, as noted above.
\end{proof}

QPP monotonicity can be further strengthened in the form of an explicit chain rule, which follows from the chain rule for classical relative entropies~\cite{cover2006book,schindler2025unification}. This chain rule is especially helpful to analyze the LOCC class (defined below) and its generalizations.

\begin{lem}[QPP chain rule]
\label{thm:qpp-chain-rule}
    Let $\rho$ be a state and \mbox{$\tau=\one/d$}. Given QPP instrument $\NN : \MM$, let $\rho_i = \MM_i(\rho)/\tr\MM_i(\rho)$ and $\tau_i = \MM_i(\tau)/\tr\MM_i(\tau)$ be the \mbox{post-$\MM$} states resulting from $\rho,\tau$, respectively. Then
    \begin{equation}
    \label{eqn:qpp-chain-rule}
        S_{\NN : \MM}(\rho) = S_\MM(\rho) - D_{\NN | \MM}(\rho \rr \tau),
    \end{equation}
    with the conditional relative entropy 
    \begin{equation}
        D_{\NN | \MM}(\rho \rr \tau) = \sum_i p_i D_{\NN^{(i)}}(\rho_i \rr \tau_i).
    \end{equation}
    Here, $\NN^{(i)} =(\NN_{j|i})_j$ is the instrument from the collection $\NN$ corresponding to outcome $i$ of $\MM$, and $D_{\NN^{(i)}}$ is the measured relative entropy.
\end{lem}

Given an initial set $\chi_0$ of available instruments, logically there is nothing stopping an experimenter from combining the measurements via QPP operations (\ie~performing them sequentially), or coarsening them via CPP data processing. Indeed, part of the usefulness of the instrument definition is that the quantum channels $\MM_i$ defining $\MM$ can (and ought to) already include experimental limitations like a state being destroyed during the experiment or decoherence during the duration of a measurement. The effect of such noise is to bring states $\rho_i,\tau_i$ appearing in the chain rule closer together, resulting in less (or no) information being extracted in subsequent rounds of measurement. (Even without noise, any \mbox{rank-1} measurement---in particular measuring in some basis---automatically renders $\rho_i = \tau_i$ for all $i$, so that no further information can be extracted; the usefulness of QPP/sequential measurement lies in the weak or coarse measurement scenario.)

Thus, so long as noise effects are included in the relevant instruments, it is fair to assume a realistic experimenter can perform any sequence of CPP and QPP steps using instruments in their set $\chi_0$. If this is the case then with any set $\chi_0$ one also has the following larger one.

\begin{defn}[PP completion]
\label{def:pp-completion}
    The set $\chi$ of all instruments that can be generated from $\chi_0$ by QPP and CPP in arbitrarily many rounds is called $\chi = {\rm PP}(\chi_0)$, the \textit{postprocessing completion} of $\chi_0$.
\end{defn}

It is sufficient to consider, without loss of generality, all the QPP (refining) rounds to occur first, followed by a final CPP (coarsening/simplifying) round. Thus ${\rm PP}(\chi_0)$ consists of all instruments $\MM = (\MM_x)_x$ with elements of the form 
\begin{equation}
\label{eqn:pp-completion-elements}
    \MM_{x} =\sum_{i_0i_1\ldots i_n} \Lambda_{x|i_0i_1\ldots i_n} \, \MM^{(n)}_{i_n|i_0i_1\ldots}  \circ \ldots  \circ \MM^{(1)}_{i_1|i_0} \circ \MM_{i_0},
\end{equation}
where each instrument involved (\eg~for each fixed $i_0$ the collection $(\MM^{(1)}_{i_1|i_0})_{i_1}$ is an instrument) is contained in $\chi_0$. Here $x$ is an arbitrary outcome label generated by the data processing round. 

While any set of measurements can be considered, those $\chi$ which are closed under postprocessing (meaning $\MM \in \chi$ implies $\Lambda \MM \in \chi$ and $\NN : \MM \in \chi$ for any $\Lambda$ and any $\NN \subset \chi$) can be easier to analyze mathematically, and are well-motivated as explained above. Among those we consider later (see below for definitions), SEP is closed under postprocessing, LO/LO* are not, and LOCC is the postprocessing completion of the LO class.

\subsubsection{Monotonicity and relative entropy of $\chi$-ness}
\label{sec:monotonicity-general}

We now turn to questions of monotonicity and invariance, which often play a central role in quantum information theoretic analysis. 

The von Neumann entropy, for example, is invariant under unitaries, meaning $S(U \rho U^\dag) = S(\rho)$ for any unitary $U$, and is monotonic under unital channels, meaning $S(\Phi(\rho)) \geq S(\rho)$ for any CPTP map~\cite{wilde2011notes} $\Phi$ with the property that $\Phi(\one) = \one$. 

Unlike von Neumann entropy, $S_M(\rho)$ (for a fixed~$M$) is \textit{not} a unitary invariant; unsurprisingly, since the state can become ``misaligned'' with $M$ under a unitary action. Rather, the correct invariance is under joint unitary rotations of both the state and measurement together, in which case $S_{UMU^\dag}(U \rho U^\dag) = S_M(\rho)$ (where in $U M U^\dag$ it is understood the rotation applies to each $M$ element). If, however, the set $\chi$ has some additional structure (\eg~if for all $M \in \chi$ also $U M U^\dag \in \chi$), then monotonicity/invariance of $E_\chi$ may hold despite failing for any particular $M\in \chi$.

We will now consider monotonicity properties that can be derived with a minimal amount of $\chi$ structure.

The class of channels (CPTP maps~\cite{wilde2011notes}) $\Phi$ under which the $\chi$-minimal entropy $S_\chi$ is a monotone are those which are both unital and $\chi$-compatible. A unital channel is one such that $\Phi(\one) = \one$. To define the latter property, one first must define the dual action of a channel on POVMs. This stems from the dual (Hilbert-Schmidt adjoint~\cite{wilde2011notes}) of a quantum channel, which is the linear map $\Phi^\dag$ defined by $\tr [Y \Phi(X) ] = \tr [\Phi^\dag(Y) X]$ for all Hermitian operators $X,Y$. When applied to a POVM this map is understood element-wise, so that $\Phi^\dag(M) = \big(\Phi^\dag(M_i) \big)_i$, and one can check that $\Phi^\dag(M)$ is indeed also a POVM so long as $\Phi$ is CPTP. Acting on a set of measurements, $\Phi^\dag(\chi) = \{\Phi^\dag(M) \, | \, M \in \chi\}$ is understood as the image. This allows the $\chi$-compatible condition to be stated as $\Phi^\dag(\chi) \subset \chi$, meaning that when $\Phi^\dag$ is applied to $M \in \chi$, the result remains inside $\chi$. In these terms we can now state the basic monotonicity results.

\begin{thm}[$S_\chi$ monotonicity]
\label{thm:channel-monotonicity-general}
    Consider a class of measurements $\chi$, and let $\Phi$ be a unital $\chi$-compatible channel (meaning $\Phi(\one)=\one$ and $\Phi^\dag(\chi) \subset \chi$). Then for any state $\rho$,
    \begin{equation}
        S_{\chi}(\Phi(\rho)) \geq S_{\chi}(\rho).
    \end{equation}
\end{thm}
\begin{proof}
    Expressing observational entropy in terms of measured relative entropy, as in~\eqref{eqn:oe-mre}, one finds
    \begin{align}
        S_{\chi}(\Phi(\rho)) 
        &= \log d - \sup_{M\in\chi} D_{M}(\Phi(\rho) \rr \one/d) \label{eq1chiunit} \\
        &= \log d - \sup_{M\in\chi} D_{M}(\Phi(\rho) \rr \Phi(\one/d)) \label{eq2chiunit} \\
        &= \log d - \sup_{M\in\chi} D_{\Phi^{\dagger}(M)}(\rho \rr \one/d) \label{eq3chiunit} \\
        &= \log d - \sup_{N\in\Phi^{\dagger}(\chi)} D_{N}(\rho \rr \one/d) \label{eq4chiunit} \\
        &\geq \log d - \sup_{N\in\chi} D_{N}(\rho \rr \one/d) = S_{\chi}(\rho). 
        \label{eqn:channel-monotonicity-general} 
    \end{align}
    The equalities follow first because $\Phi$ is unital, then by definition of the dual $\Phi^\dag$, then from the definition of the image. The inequality follows because supremizing over a superset cannot decrease the supremum value, with the $\chi$-compatible property ensuring $\chi \supset \Phi^\dag(\chi)$. It is worth noting that the fundamental inequality in this case is not relative entropy monotonicity (in contrast to many other OE properties), but rather the absorption of the map into the supremum.
\end{proof} 

A natural question is whether the gap $E_\chi(\rho)$ (a difference of two increasing monotones $S_\chi(\rho)$ and $S(\rho)$) is itself also monotonic under such maps. It is easy to come up with cases where $E_\chi$ decreases under such maps. For example, the depolarizing channel $\Phi(\rho) = \tr(\rho) \one/d$ maps any state to one with $E_\chi(\Phi(\rho))=0$, it is unital, and whenever $\chi$ contains trivial measurements it is also $\chi$-compatible. So if anything, the gap would be a decreasing monotone. However, it can also be shown that this is not the case. An example comes from the context of the separable gap $E_\sep$ that will be studied later. In that case, the relevant class of channels for $S_\chi$ monotonicity are those that are both unital and separable. In particular, the twirling by local unitaries that generates Werner states (see Subsec.~\ref{subsec:werner}) is such a map. But twirling the pure product state $\ket{0} \otimes \ket{0}$ yields the fully symmetric ($\lam = 0$) Werner state. This maps a state with $E_\sep=0$ to one with $E_\sep >0$, as seen by comparing Fig.~\ref{fig:werner-plots}. This shows that $E_\chi$ is neither an increasing nor decreasing monotone of the unital $\chi$-compatible maps. Identifying under what meaningful set of channels, if any, $E_\chi$ is monotonic, remains an open question, and is closely related to the problem of classifying ``macroscopic operations''~\cite{nagasawa2025macroscopic} associated with measurement sets $\chi$, which necessarily contain the set of $E_\chi$-monotone channels as a subset.

\smallskip

Meanwhile, another widely used class of measures are those of the form $\Delta_{\Omega}(\rho) = \inf_{\sigma \in \Omega} D(\rho \rr \sigma)$, measuring the distance of $\rho$ to some given set of states (see for example~\cite{modi2010unified} for the use of such measures to quantify quantum correlations and entanglement). We now show how $E_\chi(\rho)$ can be related to such measures by inequalities.

For a given gap $\chi$, the relevant set of states $\Omega_\chi$ for which $\Delta$ will be defined, turns out to arise from the Petz recovery of states from measurement outcomes (for review of these topics see \eg~\cite{wilde2011notes,buscemi2022observational,bai2025bayes}). 

In particular, consider $\Phi_M(\bullet) = \sum_i \tr(M_i \bullet) \ketbra{i}$, the quantum-classical channel implementing measurement~$M$. The Petz recovery map $\RR_{\Phi_M,\one/d}$ for channel $\Phi_M$ with reference state $\one/d$ can be defined as in \cite{buscemi2022observational}. The composition $\RR_{\Phi_M,\one/d} \circ \Phi_M$ results in the channel
\begin{equation}
    \PP_M(\bullet) = \sum_i \tr(M_i \bullet) \, \frac{M_i}{\tr M_i}.
\end{equation}
This channel returns from a state $\rho$ the so-called ``coarse-grained state'' $\rho_{\rm cg} = \PP_M(\rho)$, which is an estimate of the state that would be inferred from Bayesian inference on the measurement outcomes. Notably, $\PP_M$ is unital ($\PP_M(\one)=\one$), self-dual ($\PP_M^\dag = \PP_M$), and CPTP. For projective $M$, $\PP_M \circ \PP_M = \PP_M$. Furthermore, applied to a POVM one finds $\PP_M^\dag(M') = N$ where $N_j = \sum_i \Lambda_{j|i} M_i$ with $\Lambda_{j|i} = \tr(N_j M_i)/\tr(M_i)$. Therefore whenever the set $\chi$ is closed under classical post-processing (see Thm.~\ref{thm:cpp-qpp-monotonicity}) the map $\PP_M$ for $M \in \chi$ is both unital and $\chi$-compatible. 

This map is central to a powerful inequality that plays a key role in the next theorem.

\begin{lem}[Recovery inequality~\cite{buscemi2022observational}]
\label{thm:recovery-inequality}
For any POVM~$M$ and state $\rho$,
    \begin{equation}
        S(\PP_M(\rho))  \geq S_M(\rho)  \geq D(\rho \rr \PP_M(\rho)) + S(\rho).
    \end{equation}
If $M$ is projective all the equalities hold. Further, for any POVM, $S_M(\rho) = S(\rho)$ if and only if $\PP_M(\rho)=\rho$.
\end{lem}

\begin{proof}
    For completeness we comment on the proofs of these bounds, which were first proved in~\cite{buscemi2022observational}. While the lower bound inequality was proved using an asymptotic argument in~\cite{buscemi2022observational}, it can alternatively be proved using operator convexity~\cite{mosonyi2024note}, similar to what is done in Thm.~\ref{thm:sep-bound} below. Meanwhile, the upper bound is a consequence of relative entropy monotonicity applied to the Petz recovery map~\cite{Petz_recovery_2,Petz_recovery_1}. For projective $M$, equality can be observed by noting that (only in the projective case) $\PP_M(\rho)$ is the MaxEnt state giving the same outcome probabilities as~$\rho$~\cite{schindler2025unification}. This corresponds to the cross-entropy constraint  $S(\rho ; \PP_M(\rho)) = S(\PP_M(\rho))$ (consult~\cite{schindler2025unification}), which ensures $D(\rho \rr \PP_M(\rho)) = S(\PP_M(\rho)) - S(\rho)$, so the upper and lower bounds are equal. The final statement holds since relative entropy is zero only if the states are equal.
\end{proof}

With these definitions we can define the set of states
\begin{equation}
    \Omega_\chi = \{ \PP_M(\sigma) \, | \, M \in \chi, \textrm{ all states } \sigma \},
\end{equation}
which is the union of $\PP_{\chi}(\rho)$ (understood as the image of~$\chi$) over all input states. With the above lemma one can see that $\Omega_\chi$ at least contains all states such that $S_M(\rho)=S(\rho)$ for some $M \in \chi$. In terms of this set one can define the distance measure
\begin{equation}
    \Delta_{\chi}(\rho) = \inf_{\sigma \in \Omega_\chi} D(\rho \rr \sigma).
\end{equation}
As a generalization of the relative entropy of entanglement (where $\Omega$ is the set of separable states~\cite{vedral1997quantifying}) and the relative entropy of quantumness (where $\Omega$ are the set of classically correlated states~\cite{modi2010unified}), this can be called the ``relative entropy of $\chi$-ness''. This allows a statement of a general inequality relation with the gap $E_\chi$.

\begin{thm}[Relative entropy inequality]
\label{thm:relative-entropy-inequality}
    The $\chi$ entropy gap is bounded by the relative entropy of $\chi$-ness. For any $\rho,\chi$,
    \begin{equation}
        E_{\chi}(\rho) \geq \inf_{M \in \chi} D(\rho \rr \PP_M(\rho)) \geq \Delta_\chi(\rho).
    \end{equation}
\end{thm}
\begin{proof}
    The result follows directly from the observational entropy recovery inequality~\cite{buscemi2022observational}, which is Lemma~\ref{thm:recovery-inequality} above. By the lower bound, $E_\chi(\rho) \geq \inf_{M \in \chi} D(\rho \rr \PP_M(\rho))$, establishing the first inequality. But since $\PP_M(\rho) \in \Omega_\chi$, it follows that this quantity is $\geq \Delta_\chi(\rho)$, since the infimum in $\Delta_\chi$ is over a larger set.
\end{proof}

Below we will leverage the general statements of this section in the study of $E_\chi$ gaps of quantum entanglement and correlation. To set this up, we now turn to the description of OE in multipartite systems.

\section{Multipartite systems}
\label{sec:multipartite}
Henceforth we will focus on the investigation of entropy gaps in multipartite systems, particularly in the context of local measurement classes, where measurements are possible on subsystems but not necessarily on the joint global state. To address this scenario we first establish some terminology and basic properties of observational entropy in multipartite settings.

We consider an $n$-partite system $A_1 A_2 \ldots A_n$ with Hilbert space $\mathcal{H}_1 \otimes \mathcal{H}_2 \otimes \ldots \otimes \mathcal{H}_n$. The joint state of the system is denoted by $\rho$, and the reduced state of the $k$-th subsystem by $\rho_k$.

We begin by introducing a formal terminology for the simplest type of local measurement in a multipartite setting, referred to as tensor product measurements.

\begin{defn}[Tensor product measurements]
\label{eqn:product-meas}
    Let $M_1 = (M^{(1)}_{i_1})_{i_1}$, and likewise $M_k = (M^{(k)}_{i_k})_{i_k}$ for each $k=1,\ldots,n$, be POVMs on systems $A_1, \ldots, A_n$, respectively. Their tensor product is the POVM on $A_1 A_2 \ldots A_n$ defined by
    \begin{equation}
        M_1\otimes\cdots\otimes M_n = (M^{(1)}_{i_1} \otimes\cdots\otimes M^{(n)}_{i_n})_{i_1,\ldots,i_n}.
    \end{equation}
    Performing this on $\rho$ yields joint probability distribution $(p_{i_1\ldots i_n})_{i_1,\ldots,i_n}$, with $p_{i_1\ldots i_n} = \tr \big(M^{(1)}_{i_1} \otimes\cdots\otimes M^{(n)}_{i_n} \, \rho\big)$.
\end{defn}

The observational entropy of a tensor product measurement can be decomposed into a sum of marginal entropies less a mutual information term. The proof follows the same reasoning as for the case of projective measurements in~\cite{schindler2020correlation}.

\begin{prop}
\label{eqn:marginal-mutual}
    Observational entropy of tensor product measurements obeys
    \begin{equation}
    \label{oetensmeas}
    S_{M_1\otimes\cdots\otimes M_n}(\rho) = \sum_{k=1}^n S_{M_k}(\rho_k) - D(p_{i_1\ldots i_n} \rr p_{i_1}\cdots p_{i_n}),
\end{equation}
where $(p_{i_1})_{i_1},\ldots,(p_{i_n})_{i_n}$ are the marginals of the joint distribution $(p_{i_1\ldots i_n})_{i_1,\ldots,i_n}$. The relative entropy term $D(p_{i_1\ldots i_n} \rr p_{i_1}\cdots p_{i_n})$, between the joint distribution and the product of marginal distributions, is the mutual information between the subsystems.
\end{prop}

The mutual information in \eqref{oetensmeas} is non-negative, and vanishes if and only if the distributions $(p_{i_1 \ldots i_n})_{i_1,\ldots,i_n}$ and $(p_{i_1} \ldots p_{i_n})_{i_1,\ldots,i_n}$ are identical~\cite{cover2006book}. This immediately lends the properties of \textit{subadditivity},
\begin{equation}
\label{subadditivity}
    S_{M_1 \otimes \cdots \otimes M_n}(\rho) \leq \sum_{k=1}^n S_{M_k}(\rho_k),
\end{equation}
and \textit{additivity},
\begin{equation}
\label{additivity}
    S_{M_1 \otimes \cdots \otimes M_n}(\rho_1 \otimes \cdots \otimes \rho_n) = \sum_{k=1}^n S_{M_k}(\rho_k),
\end{equation}
for the observational entropy.

\smallskip

Finally, let us recall the definitions of separable states~\cite{horodecki2009entanglement} and classically correlated states~\cite{piani2011nonclassical}, which are central in the theories of quantum correlations and entanglement, and play important roles below.

\begin{defn}[Separable/entangled states]
\label{def:separable-states}
A state $\rho$ is called separable if it can be written as a convex combination of product states, in the form
\begin{equation}
    \rho = \sum_i p_i \; \rho^{(1)}_i \otimes \cdots \otimes \rho^{(n)}_i.
\end{equation}
A state that is not separable is called entangled.
\end{defn}

\begin{defn}[Classically correlated states]
\label{def:classically-correlated-states}
A state $\rho$ is called classically correlated, strictly classically correlated, or CC, if it can be written as
\begin{equation}
\label{eq67}
    \rho = \sum_{i_1,\ldots,i_n} p_{i_1\ldots i_n} \ketbra{\psi^{(1)}_{i_1}\ldots \psi^{(n)}_{i_n}}, 
\end{equation}
where $\{|\psi^{(1)}_{i_1}\rangle\}_{i_1},\ldots,\{|\psi^{(n)}_{i_n}\rangle\}_{i_n}$ are local orthonormal bases in each of the subsystems $A_1, \ldots, A_n$, and $(p_{i_1 \dots i_n})_{i_1,\dots,i_n}$ is a probability distribution. Note that $\ket{\psi^{(1)}_{i_1}\ldots \psi^{(n)}_{i_n}}$ is shorthand for $\ket{\psi^{(1)}_{i_1}} \otimes \cdots \otimes \ket{\psi^{(n)}_{i_n}}$.
\end{defn}

Separable states, \ie~convex combinations of product states, are those that can be created with local operations and classical communication~\cite{werner1989states}. Meanwhile classically correlated (CC) states are those which can be diagonalized in a product of local orthonormal bases. Every CC state is separable, and by definition separable states are not entangled. Separable states may still contain quantum correlations of the discord type, while CC states contain no quantum correlations at all~\cite{modi2010unified}.

\section{Local measurement classes and the separable measurement theorem}
\label{sec:local-measurement-classes}
We can now turn our attention to the entropy gaps of local measurement classes. For simplicity of notation, the definitions of local measurement classes will be formulated for the bipartite system $AB$; however, these definitions can be readily extended to the $n$-partite case.
\begin{itemize}
    \item \textbf{LO*} is the class of local projective measurements, that is, the class of measurements of the form $\Pi_A \otimes \Pi_B$, where $\Pi_A$ and $\Pi_B$ are projective measurements on $A$ and $B$, respectively.
    \item \textbf{LO} is the class of local measurements, that is, the class of measurements of the form $M_A \otimes M_B$, where $M_A$ and $M_B$ are POVMs on $A$ and $B$, respectively.
    \item \textbf{LOCC} is the class of measurements that can be achieved by local operations and classical communication~\cite{bennet1999nonlocality,chitambar2014locc}. This can be defined as the postprocessing completion of the LO class, as in Sec.~\ref{sec:general:cpp-qpp}. Essentially, one can perform local POVMs from the LO class in several rounds, where the POVM to be used in each round is conditioned on measurement results from previous rounds.
    \item \textbf{SEP} is the class of separable measurements~\cite{vedral1998entanglement,horodecki2009entanglement}, that is, the class of POVMs $M  = (M_i)_i$ whose elements can each be written as sums of tensor products of positive operators,
    \begin{equation}
    \label{eqn:Msep}
        M_i = \textstyle\sum_k A_{ik} \otimes B_{ik},
    \end{equation}
    with $A_{ik}, B_{ik} \geq 0$ acting on $A$ and $B$, respectively. This class subsumes all the aforementioned classes. Moreover, each POVM element is proportional to a separable state; in this sense the measurement does not utilize entanglement.
    \item We will occasionally also mention the classes \textbf{PPT} and \textbf{RCT}, which are increasingly broad relaxations of LOCC, such that $\rm SEP \subset PPT \subset RCT$, capturing certain aspects of locality. The PPT (positive partial transpose~\cite{peres1996separability,horodecki1996separability}) class consists of POVMs whose elements $M_x$ have positive partial transpose. The RCT (reduction criterion~\cite{horodecki1997reduction}) class consists of POVMs whose elements $M_x$ obey the reduction criterion $M_x \leq M_x^A \otimes \one_B$, where $M_x^A = \tr_B M_x$, and likewise in each subsystem. The set inclusions are shown in~\cite{horodecki1997reduction}. For simplicity, the main narrative below largely ignores them, but includes them where they are particularly useful or where we can remark on them in passing.
\end{itemize}
The class labels derive from the instrument viewpoint, where LO would consist of instruments achievable by local ``operations''. The star is used to indicate the restriction to projective measurements. We use the term ``local'' both broadly, to describe all these types of classes, and also sometimes specifically in reference to the LO/LO* classes; it should be clear from context which meaning is intended.

We denote the classes of POVMs within these sets of operations simply as $\chilostar, \ldots, \chirct$ (without danger of confusion), and the corresponding entropy gaps by $E_{\lostar}, \ldots, E_{\rct}$. It is straightforward to verify the inclusions
\begin{equation}
    \chilostar \subset \chilo \subset \chilo \subset \chisep,
\end{equation}
which, due to minimization over increasingly larger sets of measurements, implies the chain
\begin{equation}
\label{eqn:chain-Echi}
    E_{\lostar}(\rho) \geq E_{\lo}(\rho) \geq E_{\locc}(\rho) \geq E_{\sep}(\rho)
\end{equation}
of fundamental inequalities. These chains also extend to the PPT and RCT classes, as $\chisep \subset \chippt \subset \chirct$ also holds, as discussed above.

It is intuitively clear that SEP and LOCC are significantly more powerful than LO and LO*, but the separation between these pairs is not necessarily obvious. Later we will show that $E_\lo(\rho)=0$ if and only if $E_{\lostar}(\rho)=0$, but also give an example where $E_\lostar(\rho) > E_\lo(\rho) > 0$ is strictly unequal. Meanwhile, it is easy to find examples where $E_\lo(\rho) > E_\locc(\rho)=0$ by considering separable-but-not-classical states. Separating $E_\locc$ from $E_\sep$ is perhaps less obvious, but below we will exhibit candidate states providing a difference. 

\subsection{The separable measurement theorem}
\label{subsec:separable-meas-thm}

One expects that for the local measurement classes, the entropy gaps $E_\chi(\rho)$ are related to entanglement and quantum correlations in the state $\rho$.

A crucial step in showing this connection was taken in~\cite{schindler2020correlation}, where it was shown that $E_\lostar(\psi_{AB})$ is equal to the entanglement entropy $S_\ent(\psi_{AB}) = S(\rho_A) = S(\rho_B)$ when $\psi_{AB}$ is a bipartite pure state. But the methods used there do not extend to the LOCC or SEP classes.

To establish the connection of observational entropy to quantum correlations more broadly, it is necessary to provide a strong characterization of the limitations of the local measurement classes. This leads to the following result, which is fundamental to interpreting the local entropy gaps.

\begin{thm}[OE of separable measurements]
\label{thm:sep-bound}
Let $\rho$ be any state of a bipartite system $AB$. Then, for any $M \in \chisep$ (and in fact any $M \in \chirct$),
\begin{align}
  S_{M}(\rho) \geq \max\{S(\rho_A),S(\rho_B)\}.
\end{align}
That is, the entropy of a separable measurement is lower-bounded by the marginal von Neumann entropies.
\end{thm}

\begin{proof}
An elementary proof of this result can be obtained from the fact that the logarithm is operator monotone and operator concave~\cite{mosonyi2024note,hiai2017different}. The first step is to use operator concavity in the form of Jensen's operator inequality~\cite{HANSEN_2003}, to find
\begin{equation}
\label{eqn:sep-thm-step-1}
        \begin{split}
            S_M(\rho) &= - \sum_i p_i \log (p_i/V_i) \\
                &= - \tr \rho \left( \sum_i M_i \log (p_i/V_i) \right) \\
                & \geq - \tr \rho \log \left( \sum_i M_i p_i/V_i \right).
    \end{split}
\end{equation}
Next, since $M \in \chisep \subset \chirct$ we have that each POVM element obeys the reduction criterion $M_i \leq M_i^A \otimes \one_B$, where $M_i^A = \tr_B M_i$. Using operator monotonicity this implies, continuing from \eqref{eqn:sep-thm-step-1}, that
\begin{equation}
\label{eqn:sep-thm-step-2}
    \begin{split}
        \ldots 
         & \geq - \tr \rho \log \left( \sum_i (p_i/V_i) M_i^A \otimes \one_B \right) \\
         & = - \tr \rho \log \left( \sum_i ((p_i/V_i) M_i^A ) \otimes \one_B \right).
    \end{split}
\end{equation}
Finally, observe that $\tr_A M_i^A = \tr M_i = V_i$, so the remaining term inside the logarithm is a state $\sigma_A$. Thus due to the identity factor in $B$ one can simplify to a partial trace on system $A$. Therefore, continuing from \eqref{eqn:sep-thm-step-2}, 
\begin{equation}
    \begin{split}
       \ldots &= - \tr \rho_A \log \sigma_A \\
             &\geq -\tr \rho_A \log \rho_A \\
             &= S(\rho_A).
    \end{split}
\end{equation}
The final inequality follows from the fact that relative entropy is always non-negative. The same arguments show also $S_M(\rho)\geq S(\rho_B)$. 
\end{proof}

This result can also be proved via strengthened relative entropy monotonicity~\cite{sutter2017multivariate} in the form of the observational entropy recovery inequality~\cite{buscemi2022observational}. This approach shows that $E_\sep(\rho)$ is lower bounded by the relative entropy of entanglement, as proved in Thm.~\ref{thm:ree-and-req} below, which in turn is known to have $S(\rho_A)-S(\rho)$ as a lower bound~\cite{plenio_intr_to_ent_meas}. The above proof makes use of only the minimal necessary structure.

The multipartite version follows as an immediate corollary, since an $n$-separable measurement is in particular also separable over any bipartition of the systems into two groups.

\begin{cor}
    Let $\rho$ be any state of an $n$-partite system $A_1\otimes\cdots\otimes A_n$, and $M \in \chisep$ be fully separable with respect to the $n$-partition. Then, 
    \[
      S_M(\rho) \geq \max_{I\subseteq [n]} S(\rho_{A_I}) ,
    \]
    where $\rho_{A_I}$ is the reduced density matrix on the subsystem $A_I = \bigotimes_{i\in I} A_i$ corresponding to the index set $I$.
\end{cor}

Theorem~\ref{thm:sep-bound} seems to be a key characteristic of separable measurements within the OE framework. A simple example shows that no similar marginal bound could hold for general non-separable measurements. In fact, given an entangled pure state, say $\ket{\phi^+}= (\ket{00}+\ket{11})/\sqrt{2}$, there is always a joint measurement attaining $S_M(\ket{\psi})=0$. But no separable measurement can do better than the marginal von Neumann entropy  $S(\rho_A)= 1$ bit. More generally, this shows that information extraction by separable measurements is bounded by the inherent missing information in any subsystem. 

\subsection{Bipartite pure states}
\label{subsec:bipartite-pure-states}

The bound on entropy of separable measurements allows one to connect the entropy gaps to the entanglement entropy for any $\chi$ in the local measurement classes, when acting on bipartite pure states.

\begin{thm}[Reduction to entanglement entropy]
\label{thm:bipartite-pure-states}
Let $\psi_{AB}=\ketbra{\psi}_{AB}$ be a bipartite pure state, and denote by $S_{\ent}(\psi_{AB}) = S(\rho_A)=S(\rho_B)$ its entanglement entropy. For any $\chi$ such that $\chirct \supset \chi \supset \chilostar$, 
\begin{equation}
    E_{\chi}(\psi_{AB}) = S_\ent(\psi_{AB}).
\end{equation}
That is, for bipartite pure states all the local entropy gaps are equal to the usual entanglement entropy.
\end{thm}

\begin{proof}
Let $\ket{\psi_{AB}} \equiv \sum_k c_k \ket{a_k b_k}$ be the Schmidt decomposition of $\ket{\psi_{AB}}$. Let $M$ be a measurement in the Schmidt basis, meaning $M = M_A \otimes M_B$ where $M_A = (\ketbra{a_i})_i$ and $M_B = (\ketbra{b_j})_j$. It holds that $M \in \chilostar$, and direct evaluation shows that $S_M(\psi_{AB}) = S_\ent(\psi_{AB})$. Therefore $E_\chi(\psi_{AB}) \leq S_\ent(\psi_{AB})$, since $\chi \supset \chilostar$. However, by Thm.~\ref{thm:sep-bound}, it also holds that $E_\chi(\psi_{AB}) \geq S_\ent(\psi_{AB})$, since $\chirct \supset \chi$. Combining these results, we conclude that $E_\chi(\psi_{AB}) = S_\ent(\psi_{AB})$.
\end{proof}

This shows local $E_\chi$ gaps measure the entanglement in bipartite pure states, precisely as one would expect. 

\smallskip

Unlike for bipartite pure states---where entanglement is the only possible type of correlation---in mixed and multipartite states the landscape of classical correlations, quantum correlations, and entanglement, is far more varied. The next natural question is, how do the local $E_\chi$ gaps, for general states, fit into this broader picture.

\subsection{Entanglement, discord, and relative\protect\\ entropy measures}
\label{sec:entanglement-discord-and-re-measures}

Different types of correlations can be exhibited by quantum systems, including quantum entanglement, discord-type correlations, and classical correlations. 

These are part of a hierarchy that can be defined as follows. Any state that is not a product state is said to have correlations. Any state that is not classically correlated (Def.~\ref{def:classically-correlated-states}) is said to have quantum correlations. Any state that is not separable (Def.~\ref{def:separable-states}) is said to be entangled. Entanglement is therefore one type of quantum correlation, but not the only type; general quantum correlations are sometimes called discord or discord-type correlations. For review of this hierarchy see \eg~\cite{modi2010unified}.

These different types of correlations can be analyzed with the tools of resource theory~\cite{chitambar2019resource}. In resource theory one defines a set $\FF$ of free operations (here some set of quantum channels), and a set of free states, which for present purposes we take to be $\FF_s = \FF(\one/d)$ (the image under all free operations of the maximally mixed state). By definition free operations cannot increase the amount of resource, and resource measures are defined as functions $g(\rho)$ that are monotones, $g(\Phi(\rho)) \leq g(\rho)$ for all $\Phi \in \FF$ (measures are also sometimes assumed to have certain other features like being zero on free states). There is generally no unique choice of a resource measure, but one canonical choice is given by the relative entropy distance to the set of free states, $\Delta_\FF(\rho) = \inf_{\sigma \in \FF_s} D(\rho \rr \sigma)$, which will appear again shortly. 

When the free operations are taken to be local channels (products $\Phi_A \otimes \Phi_B$), the free states become product states ($\rho = \rho_A \otimes \rho_B$), and the relative entropy $\Delta_\FF$ becomes the total mutual information; the corresponding resource is \textit{total correlations}. When the free operations are taken to be LOCC~\cite{vidal2000entanglement} or separable (product Kraus operators $\Phi(\bullet) = \sum_k (K^A_k \otimes K^B_k) \bullet (K^A_k \otimes K^B_k)^\dag$~\cite{wilde2011notes}) channels, free states are separable states ($\rho = \sum_k \! \lam_k \, \rho^A_k \otimes \rho^B_k$), and the corresponding resource is \textit{entanglement}. The relative entropy distance to the set of separable states
\begin{equation}
    \Delta_{\textsc{ree}}(\rho) = \inf_{\sigma \in \textsc{sep}} D(\rho \rr \sigma)
\end{equation}
is a much-studied entanglement measure called the relative entropy of entanglement~\cite{vedral1997quantifying,vedral1998entanglement}. When minimising over states with positive partial transpose above, we obtain
\begin{equation}
    \Delta_{\textsc{ree,ppt}}(\rho) = \inf_{\sigma \in \textsc{ppt}} D(\rho \rr \sigma),
\end{equation}
the PPT-relative entropy of entanglement \cite{Rains:PPT}.

Unlike the previous two types of correlation, which are resources in and of themselves (within their respective theories), the \textit{discord} or \textit{quantum correlation} is better seen as a contribution to the total correlation resource---where total correlations can be broken up into classical, quantum, and other contributions~\cite{modi2010unified}. Nonetheless, there is still a clearly defined set of quantumness-free states, the classically correlated states $\rho = \sum_{ij} \rho_{ij} \ketbra{i}  \!  \otimes  \! \ketbra{j}$ (see Def.~\ref{def:classically-correlated-states}). The relative entropy distance to this set (denoted here by CC),
\begin{equation}
    \Delta_{\textsc{req}}(\rho) = \inf_{\sigma \in \textsc{cc}} D(\rho \rr \sigma),
\end{equation}
is a well-studied quantifier of quantum correlations known as the relative entropy of quantumness~\cite{piani2011nonclassical,groisman2007quantumness}. This is also called relative entropy of discord~\cite{modi2010unified} and is equivalent to measures known by many other names~\cite{bravyi2003entanglement,saitoh2008nonclassical,oppenheim2002thermodynamical,horodecki2003local,horodecki2005local,enriquez2016maximally,schindler2020correlation}.

To fit the entropy gaps into this framework, one wishes to know under which sets of channels they are monotonic. One might be tempted to speculate that the separable gap $E_{\sep}$ would be an entanglement measure, monotonic under separable operations. However, below we demonstrate (\eg~see Fig.~\ref{fig:werner-plots}) the existence of separable states $\rho \in \sep$ such that $E_{\sep}(\rho)>0$. Since any separable state can be created from $\one/d$ using LOCC or separable operations, this eliminates the possibility that $E{\chi}$ is an entanglement monotone for any $\chi \subset \chisep$.

What, then, does $E_\sep$ measure? The difference between $E_\sep$ and an entanglement measure is related to the difference between preparability and measurability. The set of states that can be \textit{prepared} by separable operations is the set of all separable states. Meanwhile, the set of states such that $E_\sep(\rho)=0$ (the so-called ``eigenseparable'' states defined below), are the set of states that can be optimally \textit{measured} by separable measurements. Similarly, the states where $E_\lostar(\rho) = 0$ are those that can be measured optimally by local projective measurement.

While not entanglement measures in the formal sense, it is clear that the local $E_\chi(\rho)$ gaps do nonetheless quantify entanglement-related quantum correlations. In fact, in the LO* case, one result of \cite{schindler2020correlation} was to show that $E_\lostar$
is precisely the relative entropy of quantumness/discord. Thus $E_\lostar$ can be understood as quantifying the amount of quantum correlations (discord) contributing to correlations in the state. In the case of $E_\sep$, a similar relation can be obtained as an inequality, showing that entanglement is at least a contribution to the $E_\sep$ gap. These two are summarized in the following theorem.

\begin{thm}[REQ and REE]
\label{thm:ree-and-req}
    For any state $\rho$ (which is $n$-partite with a fixed $n$-partition of the systems defining local/separable states/measurements), one has
    \begin{align}
        E_\lostar(\rho) &= \Delta_\req(\rho), \label{eqn:lostar-req}\\
        E_\sep(\rho) &\geq \Delta_\ree(\rho), \label{eqn:sep-ree},
    \end{align}
    relating the local entropy gaps to the relative entropies of quantumness and entanglement. Likewise,
    \begin{equation}
        E_\ppt(\rho) \geq \Delta_{\ree,\ppt}(\rho). \label{eqn:ppt-ree}
    \end{equation}
\end{thm}
\begin{proof}
    The $E_\lostar$ equality was proved in~\cite{schindler2020correlation}. For $E_\sep$ we make use of Thm.~\ref{thm:relative-entropy-inequality} above. For any $M \in \chisep$ the coarse-grained state $\PP_M(\rho)$ is necessarily a separable state, as it is a convex combination of separable POVM elements. Therefore $\Omega_\sep$ is a subset of the separable states, implying that $\inf_{\sigma \in \sep} D(\rho \rr \sigma) \leq \inf_{\sigma \in \Omega_\sep} D(\rho \rr \sigma)$, which shows that $\Delta_\chi$ in Thm.~\ref{thm:relative-entropy-inequality} is $\geq \Delta_\ree$.
    The proof for $\chi=\text{PPT}$ is similar.
\end{proof}

As discussed below Thm.~\ref{thm:channel-monotonicity-general}, it is currently not known under which channels the gaps $E_\chi$ are monotone. However, with respect to $S_\chi$ some results are known.

\begin{thm}[Channel monotonicity]
\label{thm:channel-monotonicity-local-separable}
With unital, local, and separable channels defined as in the text above, for all states $\rho$,
   \begin{enumerate}[(i)]
       \item $S_\lo\big(\Phi(\rho)\big) \geq S_\lo(\rho)$ for any $\Phi$ which is both local and unital,

       \item $S_\sep\big(\Phi(\rho)\big) \geq S_\sep(\rho)$ for any $\Phi$ which is both separable and unital.
   \end{enumerate}
\end{thm}

\begin{proof}
    This follows from the general Thm.~\ref{thm:channel-monotonicity-general}. For SEP, since $\Phi$ has separable Kraus operators, it follows that $\Phi^\dag(\chisep) \subset \chisep$. For LO, since $\Phi$ must be a product of unital channels, $\Phi_A^\dag(M_A)$ is a local POVM in system $A$, and so on in each subsystem, so $\Phi^\dag(\chilo) \subset \chilo$.
\end{proof}

The dual presence of both unital and local/separable conditions in the monotonicity theorem suggests the natural setting for these gaps is a joint resource theory of purity (the resource associated with unital maps) and correlations/entanglement (the resources associated with local/separable maps). Interactions between purity and quantum correlations have been explored \mbox{previously~\cite{zyczkowski1998volume,piani2011nonclassical}}, but the full scope of such a theory in the resource context remains for further study.

\subsection{The multipartite picture}
\label{sec:multipartite-picture}

A convenient feature of how the entropy gaps quantify quantum correlations is that the value of $E_\chi(\rho)$ is on an absolute and physically meaningful scale, comparable to other entropies of the system.

One upshot of this is that the gaps $E_\chi$ in multipartite systems can be quantitatively compared across different types of partition. 

For this purpose we can introduce a simple notation to denote different partitions. For instance given a 4-partite system $ABCD$ one can represent various partitions by $AB|CD$, $AC|BD$, $A|B|C|D$, $ACD|B$, and so on. Throughout the paper we primarily assume the partition is obvious from context, and only specify partitions explicitly when it is relevant.

One simple yet important property is the monotonicity of the gaps under refinement of the partition.

\begin{prop}[Partition monotonicity]
\label{prop:part-mon}
    For any $\chi$ in the local classes, the gap is a monotonic function under adding an additional partition. For example, $E_\chi^{A|BC}(\rho) \leq E_\chi^{A|B|C}(\rho)$ for all states $\rho$.
\end{prop}

\begin{proof}
    Any $M$ which is local with respect to $A|B|C$ is in particular also local with respect to $A|BC$, so that $\chi^{A|B|C} \subset \chi^{A|BC}$ for whichever of the classes is relevant. The infimum over a subset gives a $\geq$ value.
\end{proof}

The fact that gaps can be quantitatively compared across different types of partitions makes them useful for analyzing the different ways quantum correlations can be distributed throughout a multipartite network~\cite{bengtsson2016multipartite,walter2016multipartite}, such as the genuineness~\cite{entanglement2009guhne} and robustness~\cite{dur2000three,neven2018robustness} of multipartite correlations. An example comparing genuineness and robustness in 4-partite GHZ, $W$, and two-Bell-pair states is given in Fig.~\ref{fig:multipartite-genuine-robust}, in the later section on examples.

\smallskip
This completes the main theoretical developments of the paper. We now turn to more phenomenological aspects of the theory. In the following sections we establish separations and equivalences between the different gaps, demonstrate some theorems which are useful for evaluating or bounding the gaps, evaluate certain examples, and discuss the problem of identifying the optimal measurement from the given classes.

\section{LO versus LO*}
\label{sec:lovslostar}

Among all the classes, the LO* gaps are by far the easiest to evaluate either analytically or numerically. This is because, due to CPP monotonicity of the entropy under refinement of POVMs, the problem of choosing optimal projective measurements reduces to that of finding an optimal measurement basis, so the LO* case reduces to an optimization over local unitaries. This has allowed $E_\lostar$ gaps to be evaluated for many states, in studies of various equivalent quantities~\cite{bravyi2003entanglement,saitoh2008nonclassical,modi2010unified,piani2011nonclassical,enriquez2016maximally,schindler2020correlation}. The problem of choosing optimal POVMs is much more involved, as the space of possibilities is more complex, and generally requires additional simplifying structure to make progress.

It is already clear that states $\rho$ with $E_\lostar(\rho)=0$ are precisely the classically correlated states, from the equivalence $E_\lostar=\Delta_\req$ in \eqref{eqn:lostar-req}. For classically correlated states, a $\rho$ eigenbasis measurement is already itself in LO*, and is guaranteed to reveal minimum entropy. Meanwhile, quintessential examples of states with nonzero $E_\lostar$ gaps include  $\rho = (\ketbra{00} + \ketbra{1+})/2$, and $\ket{\phi^+} = (\ket{00}+\ket{11})/\sqrt{2}$. The latter, a Bell pair, is entangled, and has $E_\lostar = 1$~bit in accordance with its entanglement entropy. The former, with $E_\lostar = 0.5$~bits, is a separable state, so its gap is due entirely to discord-type correlations. 

Beyond projective measurements, observers in the LO class can perform local POVMs on each subsystem. Given that $E_{\lo} \leq E_{\lostar}$ due to containment of the set over which the infimum occurs, classically correlated states must satisfy $E_{\lo}(\rho) = 0$. A priori, however, there could be other states for which $E_\lo=0$ but $E_\lostar$~not. Nevertheless, it is reasonable to expect $E_{\lo}$ to be nonzero whenever quantum correlations are present, as LO measurements—though more general than projective ones—are still performed locally by the observers. There is no intrinsic reason to assume that LO measurements do not detect such correlations. Formally, this implies that if $E_{\lo}(\rho) = 0$, then $\rho$ must be classically correlated. This is the assertion established in the next theorem, showing that $E_\lo(\rho) = 0$ is equivalent to $E_{\lostar}(\rho)=0$.

\begin{thm}
\label{thm:LO-LOstar-zero-equivalence}
    $E_{\lo}(\rho)=0$ if and only if $E_{\lostar}(\rho)=0$ if and only if $\rho$ is a strictly classically correlated state.
\end{thm}
\begin{proof}
    It was shown in \cite{schindler2020correlation} that $E_\lostar(\rho) = 0$ if and only if $\rho$ is strictly classically correlated, and from \eqref{eqn:chain-Echi} it is clear that $E_\lostar(\rho) = 0$ implies $E_\lo(\rho) = 0$. We now show that $E_\lo(\rho) = 0$ implies $E_\lostar(\rho) = 0$. For simplicity the proof is stated for the bipartite case, but the same proof can easily be modified to apply to any multipartition.
    
    Suppose $M = (M_i)_{i \in I}$ and  $N = (N_j)_{j \in J}$ are local POVMs such that $S_{M \otimes N}(\rho)=S(\rho)$. Their product is $M \otimes N = (M_i \otimes N_j)_{i,j}$ as defined in \eqref{eqn:product-meas}. From Lemma~\ref{lemma:optimal-meas}, $M_i \otimes N_j = \Pi_{k(i,j)} (M_i \otimes N_j)\Pi_{k(i,j)}$ for some $k(i,j)$, where $k$ label eigenprojectors in $\rho = \sum_k \lam_k \Pi_k$, the spectral decomposition of $\rho$. Moreover, from this it follows that any overlapping (meaning $(M_i \otimes N_j)(M_{i'} \otimes N_{j'}) \neq 0$) elements must obey $k(i,j) = k(i',j')$ and thus lie in the same $\rho$ eigenspace. We will use this fact to show that the elements $M_i$ can be grouped into subsets so as to form a local projective measurement, and likewise the $N_j$, such that the projective versions give the same entropy.
    
    To do so we define a ``minimal projective coarsening'' of a POVM, in particular by decomposing $M$ into subsets using an equivalence relation defined as follows. Say $M_i$ and $M_{i'}$ overlap if $M_i M_{i'} \neq 0$, and say $i \sim i'$ if there exists any sequence $M_{i_0}, M_{i_1}, \ldots M_{i_K}$ of elements such that $i_0 = i$, $i_K = i'$, and each element overlaps the subsequent one in the sequence. Let $m$ be a label for the equivalence classes, and let $I_m \subset I$ be the subset of $M$ outcomes lying in that equivalence class. Equivalently, these subsets $I_m$ can be constructed by starting from any single element, then repeatedly adding in any elements that overlap some element of the existing subset, and proceeding iteratively until termination. The result is a decomposition of $I$ into disjoint subsets $I_m$ such that any two $M_i, M_{i'}$ in the same $ I_m$ are linked by a sequence of overlapping elements, and any two $M_i, M_{i'}$ in different $I_m,I_{m'}$ must be non-overlapping. 

    Now define operators $P^A_m = \sum_{i \in I_m} M_i$ that are sums over the POVM elements in each class. These are positive semidefinite, and since $\sum_m P^A_m = \sum_i M_i = \one$, also complete. By construction the elements of different~$I_m$ are non-overlapping, thus also $P^A_m P^A_{m'} = 0$ for $m \neq m'$. With this fact and completeness, it also follows that $P^A_m = P^A_m  \one  = (P^A_m)^2$. Together, this shows $P^A = (P^A_m)_m$ is a local projective measurement in system $A$. Likewise, construct $P^B = (P^B_n)_n$ in system $B$. 

    Now observe that for any $M_i \otimes N_j$ and $M_{i'} \otimes N_{j'}$ such that $i,i' \in I_m$ and $j,j' \in J_n$ (the system $B$ equivalent of~$I_m$), there exists a chain of $(i_k,j_k)$ pairs connecting them such that each element of the chain overlaps the next one. Therefore $k(i,j)=k(i',j')$, since the equality must hold at each step along the chain, using the property of overlapping elements mentioned earlier. This implies that each projector $P^A_m \otimes P^B_n$ has a particular $k = k(m,n)$ for which $\Pi_{k} (P^A_m \otimes P^B_n) \Pi_{k} = P^A_m \otimes P^B_n$, since each term in the sum expansion has the same $k(i,j)$. This is precisely the condition required by Lemma~\ref{lemma:optimal-meas} to show that $S_{P^A \otimes P^B}(\rho) = S(\rho)$. Since $P^A \otimes P^B$ is an LO* measurement giving an entropy equal to $S(\rho)$, it follows that $E_{\lostar}(\rho) = 0$.

    This establishes the theorem in finite dimensions, where the set of POVMs in $\chilo$ is compact (see~\cite{Compact_POVM_finite_d}) and therefore the infimum is realized as a minimum. In infinite dimensions the same proof ensures $E_{\lo}(\rho) = 0$ implies $E_{\lostar}(\rho) = 0$ if the LO infimum is achieved by some $M$ as a minimum. The possibility that the infimum is not realized must be treated separately and is for now left open.
\end{proof}

So for merely detecting the presence of quantum correlations, $E_{\lo}$ and $E_{\lostar}$ are equivalent.

This raises the question of whether local POVMs are ever, in fact, more powerful for minimizing entropy than local projective measurements? The answer is yes; despite being equivalent in the zero case, the $E_\lo$ and $E_\lostar$ gaps are not equal in general.

To demonstrate this we first establish some facts about classical-quantum states, where the problem of optimizing LO/LO* measurements can be simplified.

A classical-quantum (CQ) state~\cite{wilde2011notes} is defined as a state of the form
\begin{equation}
\label{eqn:CQ-state}
    \rho = \sum_k \lambda_k \, \ketbra{k} \otimes \rho_k,
\end{equation}
where $\{ \ket{k} \}_k$ is an orthonormal basis in one subsystem, $\rho_k$ are arbitrary states in the other subsystem, and $\{\lambda_k\}_k$ is a probability distribution. For CQ states it can be shown that the optimal measurement on the ``classical'' side must be a measurement in this same $\{ \ket{k} \}_k$ basis.

\begin{thm}
\label{thm:CQ-state_LO-LOstar}
    Consider a CQ state $\rho = \sum_k \! \lam_k \ketbra{k} \otimes \rho_k$. Let $M,N$ be any POVMs. Let $\Pi_{\cb} = (\ketbra{k})_k$ denote the measurement in the classical basis. Then 
    \begin{equation}
        S_{M\otimes N}(\rho) \geq S_{\Pi_\cb \otimes N} (\rho).
    \end{equation}
\end{thm}
\begin{proof}
    The outcome distribution for $\Pi_\cb \otimes N$ is given by $p^{\cb}_{kj} \equiv \tr [(\ketbra{k} \! \otimes \! N_j )\rho] = \lam_k \tr N_j \rho_k$. For $M \otimes N$ it is $p_{ij} = \tr [(M_i \! \otimes \! N_j )\rho] = \sum_k \bra{k} M_i \ket{k} \lambda_k \tr N_j \rho_k$, which is $p_{ij} = \sum_{k'j'} \Lambda_{ij|k'j'} p_{k'j'}^\cb$ with $\Lambda_{ij|k'j'} = \bra{k'} M_i \ket{k'} \delta_{jj'}$. In other words, $p = \Lambda p^\cb$ are related by a CPP channel. If one applies these two measurements to the state $\one/d$, then similarly $q = \Lambda q^{\cb}$ with the same $\Lambda$. Thus using \eqref{eqn:oe-mre}, by monotonicity of relative entropy under classical channels~\cite{wilde2011notes}, one has $S_{M\otimes N}(\rho) = \log d - D(\Lambda p^\cb \rr \Lambda q^\cb) \geq \log d - D(p^\cb \rr q^\cb) = S_{\Pi_\cb \otimes N} (\rho)$. In summary, when applied to the CQ state, it is as if $M \otimes N$ were coarser than $\Pi_\cb \otimes N$ in the CPP sense (even though the CPP relation does not generally hold at the POVM level).
\end{proof}

Therefore, for CQ states the problem of finding optimal LO/LO* measurements reduces to finding the optimal choice of $N$ in the Q side of the system. Meanwhile, with help of \eqref{eqn:marginal-mutual}, the associated entropy reduces to
\begin{equation}
\label{eqn:CQ-ensemble-mutual-info}
    S_{\Pi_\cb \otimes N}(\rho) = S(\rho_A) + S_N(\rho_B) - I(N : \EE),
\end{equation}
where $I(N:\EE)$ is the mutual information between the POVM $N$ and the ensemble $\EE = \{\lam_k, \rho_k\}$, defined by $I(N : \EE) = D(p_{jk} \rr p_j p_k)$ with $p_{jk} = \lam_k \tr N_j \rho_k$ and $p_j, p_k$ its marginals, and $\rho_A$, $\rho_B$ are the reduced states in each subsystem. This form is especially useful due to the presence of extensive studies on optimizing $I(N:\EE)$ in the context of accessible information and Holevo's bound~\cite{wilde2011notes}. For symmetric ensembles such that $\sum_k \lam_k \rho_k = \one/d$, the OE optimization problem reduces to the problem of optimizing $I(N:\EE)$.

\smallskip

\begin{figure}[t]
    \centering
    \includegraphics[width=.8\linewidth]{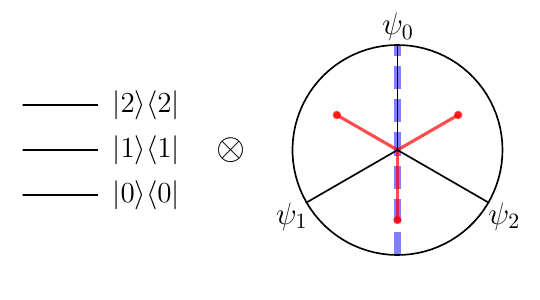}
    \caption{An example for which $E_\lo(\rho) \neq E_\lostar(\rho)$ is given by classical-quantum ``trine'' state  $\rho = \frac{1}{3} \sum_{k=0}^{2} \ketbra{k} \otimes \ketbra{\psi_k}$. Optimal LO/LO* measurements take the form $\Pi_\cb \otimes N$ where $\Pi_\cb$ measures system $A$ in the $\{\ket{0},\ket{1},\ket{2}\}$ basis. System $B$, a qubit, is depicted on the $xy$ plane in the Bloch sphere. For LO the optimal $N$ is the anti-trine POVM (red), while for LO* measuring on an aligned basis (blue dashed) is optimal.}
    \label{fig:trine-illustration}
\end{figure}

We now give an example where $E_\lo(\rho) \neq E_\lostar(\rho)$, demonstrating inequivalence between the classes. The example is a classical-quantum \textit{trine} state~\cite{trine_state}, illustrated in Fig.~\ref{fig:trine-illustration}. With system $A$ a qubit and $B$ a 3-level system, the CQ trine state is defined by
\begin{equation}
    \rho_{AB} = \frac{1}{3} \sum_{k=0}^{2} \ketbra{k} \otimes \ketbra{\psi_k} ,
\end{equation}
where $\ket{\psi_j} = \frac{1}{\sqrt{2}} (\ket{0} + e^{2\pi i k / 3} \ket{1})$ are equidistant states on the $xy$-plane of the Bloch sphere and  $\{\ket{k} \}_k$ is an orthonormal basis. 

Observe that $\sum_k \frac{1}{3}\ketbra{\psi_k} = \one_B/2$ and that since $\rho$ is an equal mixture of orthogonal states, $S(\rho) = \log 3$. Thus with \eqref{eqn:CQ-ensemble-mutual-info} one obtains that
\begin{equation}
    S_{\Pi_\cb \otimes N}(\rho_{AB}) - S(\rho_{AB}) = 1 - I(N:\EE).
\end{equation}
According to~\cite{anti_trine_POVM}, the POVM that maximizes $I(N:\EE)$ for the trine ensemble is the so-called ``anti-trine'' measurement $N = (\frac{2}{3}\ketbra{\psi_k^\perp})_k$, where $\ket{\psi_k^\perp}$ are the states orthogonal to $\ket{\psi_k}$. This yields LO gap
\begin{equation}
    E_{\lo}(\rho_{AB}) = 2 - \log 3 \approx 0.415. 
\end{equation}
For LO*, the gap is obtained by performing the measurement $(\ketbra{k})_k$ on system $A$ (see Thm.~\ref{thm:CQ-state_LO-LOstar}). The task then reduces to finding the measurement $N$ on $B$ that minimizes the OE. Consider $N$ as a qubit measurement along an arbitrary axis $\hat{n} \equiv (n_x, n_y, n_z)$ in the Bloch sphere. We find that the OE reaches its minimum when $\hat{n}$ lies in the $xy$-plane. By restricting our analysis to these axes, the OE depends solely on $n_x$. The minimum is attained for $n_x = 1$, meaning that the optimal measurement on $B$ is along the $x$ axis, as illustrated in Fig.~\ref{fig:trine-illustration}. This leads to the gap
\begin{equation}
\label{elo*trine}
    E_{\lostar}(\rho_{AB}) = \frac{4}{3} - \frac{1}{2} \log 3 \approx 0.541. 
\end{equation}
Thus $E_\lo(\rho) < E_\lostar(\rho)$ in this example.

\section{LOCC}
\label{sec:locc}
The LOCC (local operations and classical communication~\cite{horodecki2009entanglement}) measurement class considers observers who can not only do local POVMs, but can choose them conditionally based on earlier local measurement outcomes.

The quintessential example of a state that can be measured well by LOCC, but not LO/LO*, is the separable, but not classical, state $\rho = (\ketbra{00} + \ketbra{1+})/2$. Earlier it was mentioned that $E_\lostar(\rho)=0.5$~bits for this state, and by Thm.~\ref{thm:LO-LOstar-zero-equivalence} this also implies that $E_\lo(\rho)>0$. Meanwhile, for this state, $E_\locc(\rho)=0$.

To measure this state optimally using LOCC, first one measures system $A$ in the $\{\ket{0},\ket{1}\}$ basis. Then, if outcome $0$ is obtained, system $B$ is measured in the $\{\ket{0},\ket{1}\}$ basis. Otherwise, if outcome $1$ is obtained, system $B$ is measured in the $\{\ket{+},\ket{-}\}$ basis. This procedure is equivalent to measuring in the global basis $\{ \ket{00}, \ket{01}, \ket{1-}, \ket{1+}\}$, which is a $\rho$ eigenbasis, and therefore reveals $S_M(\rho)=S(\rho)$ for this state. This is an example of a ``one-way'' LOCC protocol, where each system is only measured once. It can be seen clearly from this example why LOCC is indeed a far more powerful measurement class than LO or LO*.

Formally, here we have defined LOCC as the postprocessing completion (in the sense of Def.~\ref{def:pp-completion}) of the set of local instruments, which is equivalent to the usual definitions in the literature~\cite{chitambar2014locc}. Therefore the set of LOCC instruments consists of all $\MM$ with elements of the form~\eqref{eqn:pp-completion-elements}, with each instrument in the chain being a local (tensor product) instrument. 

When searching for optimal LOCC measurements it is clear that the final CPP step appearing in~\eqref{eqn:pp-completion-elements} can only make the entropy larger, by the earlier monotonicity theorems. Therefore, for evaluating $E_\locc$, it is sufficient to consider the minimal entropy that can be extracted via local instruments and QPP.

It is useful to think of LOCC measurement as occurring in several QPP rounds, where in each round a local instrument is performed on exactly one subsystem (meaning one performs an instrument of the form $\MM_{\rm sys} \otimes \one_{\rm else}$). A common question in the context of state discrimination and related problems is: how much advantage is gained by performing many rounds instead of just a few~\cite{kleinmann2011asymptotically}? In the entropy context, the QPP chain rule~\eqref{eqn:qpp-chain-rule} can be used to bound the utility of additional rounds.

Consider the case of an $n$-partite system $A_1 \ldots A_n$. Obviously it is advantageous to at least perform one measurement in each subsystem. After these first $n$ rounds, denote by $\MM^n$ the instrument that has effectively been performed. This instrument has elements $\MM^n = (\MM_{i_1, \ldots, i_n})_{i_1,\ldots,i_n}$ of the form
\begin{equation}
    \MM_{i_1, \ldots, i_n} = \AAA^1_{i_1} \otimes  \AAA^2_{i_2|i_1} \otimes \ldots \otimes \AAA^n_{i_n|i_{n-1} \ldots i_1},
\end{equation}
where each $\AAA^k$ (for fixed conditional indices) represents a local instrument on the $k$th subsystem. If one subsequently performs any additional instrument $\MM'$, the QPP chain rule bounds $\Delta S = S_{\MM_n}(\rho) - S_{\MM':\MM_n}(\rho)$, the additional extracted entropy, by (with the shorthand notation $i \equiv (i_1,\ldots,i_n)$)
\begin{equation}
    \Delta S \leq \sum_{i}  p_{i} \,  D_{\MM'}(\rho_{i} \rr  \tau_{i}) \leq \sum_{i}  p_{i}  \, D(\rho_{i} \rr  \tau_{i})
\end{equation}
where $p_{i} = \tr \MM_{i}(\rho)$, $\rho_{i} = \MM_{i}(\rho) / p_{i}$, $\tau_{i} = \MM_{i}(\one/d)/q_i$, and $q_{i} = \tr \MM_{i}(\one/d)$. Furthermore, if the $(n+1)$-th round consists of a local instrument $\BBB^k$ on system $A_k$, this is further simplified to
\begin{equation}
    \Delta S \leq \sum_{i}  p_{i} \,  D_{\BBB^k}(\rho_{i}^{A_k} \rr  \tau_{i}^{A_k}) ,
\end{equation}
depending only on reduced conditional states $\rho_{i}^{A_k}$, $\tau_{i}^{A_k}$  in the subsystems. These immediately impose bounds on additional utility based on the distinguishability of the post-measurement state $\rho_{i}$ and the post-measurement state $\tau_{i}$ that would be obtained from the maximally mixed state, after the first $n$ rounds. If the initial measurements $\AAA^k$ are sufficiently fine-grained, for example if a complete basis is measured in each system, then $\rho_i=\tau_i$ and there is no purpose to further rounds. Only if initially weak measurements are performed in the first round is there a reason to continue the protocol.

We now return to the one-way case, where only one instrument is to be performed in each system, and focus for simplicity on the bipartite case of a system $AB$. First $\AAA = (\AAA_i)_i$ is performed on $A$, then $\BBB^{(i)}=(\BBB_{j|i})_j$ is conditionally performed on~$B$. To be explicit, this means that 
$\MM_A = (\AAA_i \otimes \one_B)_i$ and $\MM_B^{(i)}=(\one_A \otimes \BBB_{j|i})_j$ are performed on the joint global system. One useful observation is that, after performing $\MM_A$, the conditional reduced state in $B$ remains $\tau^B_i = \one_B/d_B$ regardless of the initial outcome. Therefore, from the QPP chain rule one can find that
\begin{equation}
    S_{\MM_B : \MM_A}(\rho) = S_{\AAA}(\rho^A) + \sum_i p_i S_{\BBB^{(i)}}(\rho_i^B),
\end{equation}
where $p_i = \tr \AAA_i(\rho^A)$ and $\rho_i^B = \tr_A [(\AAA_i \! \otimes  \! \one_B)(\rho)]$, with reduced state $\rho^A = \tr_B \rho$. This can also be extended to the $n$-partite case. This shows that during the initial one-way phase of LOCC, one optimization strategy is to sequentially optimize conditional entropies in each system. However, the larger questions of when a one-way strategy might be sufficient, and which order to optimally measure, are more complicated topics to resolve.

Regarding the question of for what states \mbox{$E_{\locc}(\rho)=0$}, we defer for now the question to the separable case, which is easier to analyze mathematically. Any state satisfying $E_\locc(\rho)=0$ necessarily also satisfies $E_\sep(\rho)=0$, and so at least must obey the property of eigenseparability laid out in the following section.

It remains to consider separation of $E_\locc$ and $E_\sep$. We establish this by constructing a state for which $E_\sep(\rho)=0$ and $E_\locc(\rho)>0$, defined as follows: let $\ket{\alpha_i}\otimes\ket{\beta_i}$, $i=1,\ldots,9$ be the orthonormal basis of $\mathbb{C}^3\otimes\mathbb{C}^3$ defined in \cite{bennet1999nonlocality}, where it is shown that the measurement $M=(M_i:=\ketbra{\alpha_i}\otimes\ketbra{\beta_i}:i=1,\ldots,9)$ is not implementable by LOCC (while evidently it is in SEP). In fact, it is not even in the closure $\overline{\text{LOCC}}$, meaning that there is a $\delta>0$ such that every LOCC-implementable POVM $(M_i')$ is $\delta$-far from $(M_i)$: $\sum_{i=1}^9 \|M_i-M_i'\| \geq \delta$. Now consider $\rho = \sum_{i=1}^9 p_i M_i$, with a probability distribution of positive and pairwise distinct values $p_i>0$, $p_i\neq p_j$ for all $i\neq j$. Clearly, $E_\sep(\rho) = 0$, since the eigenprojections are precisely the $M_i=\ketbra{\alpha_i}\otimes\ketbra{\beta_i}$ (by way of the eigenvalues all being different, this is unique), which are tensor products (see Lemma~\ref{lemma:optimal-meas}, as well as Thm.~\ref{th:eigensep} below). By the same Lemma \ref{lemma:optimal-meas}, $E_\locc(\rho)=0$ would imply that a fine-graining of $M$---and hence $M$ itself---is in $\overline{\text{LOCC}}$ (since $M \to M_\rho$ in the sense of Lemma \ref{lemma:optimal-meas}d), but that is exactly what the main result of \cite{bennet1999nonlocality} excludes; thus, necessarily $E_\locc(\rho) > 0$.

\section{SEP}
\label{sec:SEP}
In the information theoretic context where active observers attempt to extract information about a nonlocally distributed state, LOCC is the most realistic measurement class to describe the scenario. The main importance of the SEP class is that it is the tightest relaxation of LOCC that is mathematically far simpler to analyze~\cite{bennet1999nonlocality}. Meanwhile, in the context of physics and statistical mechanics, the LO* class is the one that seems to arise most naturally, as thermodynamic entropies are often related to tensor products of local observables (such as measurements  $M_{H_S} \otimes M_{H_B}$ of local Hamiltonians in a system and bath)~\cite{schindler2025unification}. In this way, LO* and SEP form the boundaries of the local classes that are most relevant in informational and physical contexts.

The utility of SEP is the mathematically simple form of its definition, which is that POVM elements for SEP measurements have the form
\begin{equation}
\label{eqn:-sep-def}
    M_k = \sum_{l} A^1_{k,l} \otimes \ldots \otimes A^n_{k,l}, 
\end{equation}
where all $A^m_{k,l} \geq 0$. In other words, the POVM elements are proportional to separable states.

Some regularity properties are worthy of note, which may help in the problem of optimization: SEP is a convex set, is closed under both CPP and QPP, is closed under disjoint (called ``flagged'' in~\cite{schindler2023continuity}) convex combinations, and its general elements can be given in a closed form. None of the smaller classes above have all of these properties.

Due to monotonicity of the entropy under CPP, the search for optimal separable measurements can be reduced to the case where each POVM element is proportional to a product pure state. To see this, choose an eigenbasis for each local operator $A^m_l$ in \eqref{eqn:-sep-def} above, so that $M_k$ becomes a sum of rank-1 elements. Separating this sum into separate outcome labels is a CPP refinement, and can only decrease the entropy. Therefore, an optimal SEP measurement must have the form $M = (v_k \ketbra{a_k} \otimes \ketbra{b_k} \otimes \ldots)_k$, where $v_k \geq 0$ and $\ket{a_k},\ket{b_k},\ldots$ are arbitrary local unit vectors.

The relation of $E_\sep$ to quantum entanglement theory is discussed at length in Sec.~\ref{sec:entanglement-discord-and-re-measures}. One observation made there is that rather than being an entanglement measure, $E_\sep$ is instead related to a combination of entanglement and purity. Indeed, as we will shortly see, $E_\sep$ cannot be strictly an entanglement measure, as there exist separable states $\rho$ for which $E_\sep(\rho) >0$; not all states that can be prepared by separable operations can be optimally measured by separable measurements. Instead, the class of states for which $E_\sep(\rho)=0$ is characterized by the following theorem.

\begin{defn}[Eigenseparable]
A state $\rho$ is called \emph{eigenseparable} if all projectors onto the eigenspaces of $\rho$, including the zero eigenspace (kernel of $\rho$), are separable. 
    
More formally, let $\rho$ have the spectral decomposition $\rho = \sum_k \lam_k \Pi_k$, where $\lam_k > 0$ are distinct positive eigenvalues. The zero eigenspace projector is $\Pi' = \one - \sum_k \Pi_k$. If $\Pi'$ and all $\Pi_k$ are separable projectors, then $\rho$ is eigenseparable.  
\end{defn}

\begin{thm}[Zero separable gap]
\label{th:eigensep}
$E_\sep(\rho)=0$ if and only if $\rho$ is eigenseparable.
\end{thm}

\begin{proof}
This is an immediate consequence of Lemma~\ref{lemma:optimal-meas}. If $\rho$ is eigenseparable, its eigenprojectors define a measurement in SEP. Meanwhile, if there exists an $M\in \sep$ such that $S_M(\rho)=S(\rho)$ then by Lemma~\ref{lemma:optimal-meas}, each $\rho$ eigenprojector is a sum of POVM elements, which since $M \in \sep$ is separable. This establishes the theorem in finite dimensions, where SEP is compact~\cite{Compact_POVM_finite_d}.
\end{proof}

The above proof works beyond finite dimension, whenever the infimum defining $E_\sep$ is obtained as a minimum. Infinite-dimensional cases where the infimum is not realized must be treated by separate methods. 

\begin{figure*}[t]
    \centering
    \begin{tabular}{ll}
         (a) & (b) \\
         \includegraphics[width=.45\textwidth]{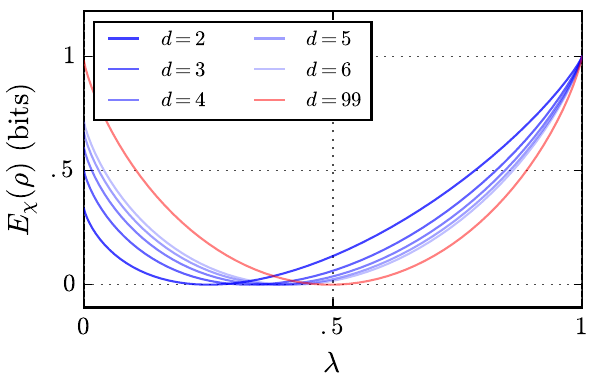}
         &
         \includegraphics[width=.45\textwidth]{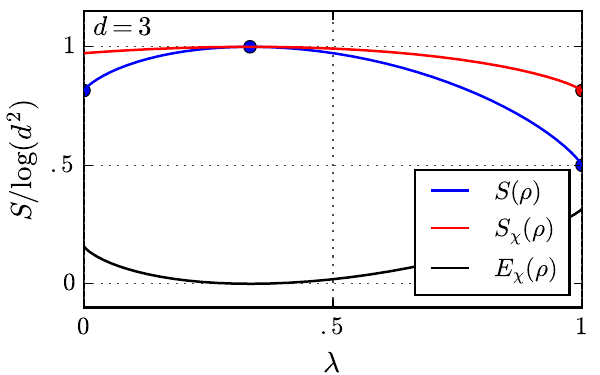}
    \end{tabular}
    \caption{For Werner states $E_\ppt(\rho_\lam)=E_\lostar(\rho_\lam)$, so all gaps in the local classes are equal. Here we plot $E_\chi(\rho_\lam)$ for any $\chippt \supset \chi \supset \chilostar$. Panel (a) shows the gap $E_\chi(\rho_\lam)$ as a function of $\lam$ for various Hilbert space dimensions. Panel (b) shows how $E_\chi(\rho_\lam)$ arises from a combination of $S_M(\rho_\lam)$ and $S(\rho_\lam)$ for the case $d=3$.   For $0 \leq \lam \leq 0.5$ the Werner states are separable, otherwise they are entangled. However, the only Werner state with $E_\chi(\rho_\lam) = 0$ is the maximally mixed state. All other Werner states, even the separable ones, exhibit a nonzero gap.}
    \label{fig:werner-plots}
\end{figure*}

\medskip
An example of a separable state where $E_\sep(\rho)>0$ is given by the Werner states~\cite{werner1989states}. In Subsec.~\ref{subsec:werner} in the next section we find that on all Werner states, $E_\sep(\rho)=E_\lostar(\rho)$, meaning all the local classes have equal gaps, and it is seen in Fig.~\ref{fig:werner-plots} that for all Werner states except the maximally mixed state, this gap is $>0$. Since there are separable Werner states, different from the maximally mixed state, these provide examples of the desired type. Furthermore, as we will explicitly show in Subsec.~\ref{subsec:werner}, no Werner state is eigenseparable, except in some specific case.

Other examples of this type can be found by mixing entangled states with noise. If $\rho$ is entangled it is not eigenseparable, and mixing $\rho$ with $\one/d$ does not change its eigenprojectors (except in discrete cases where degeneracies are introduced). However, mixing with a sufficient amount of $\one/d$ always yields a separable state~\cite{gurvits2002largest}. This gives a continuum of examples of this type.

Every eigenseparable state is separable. However, we have just seen that not every separable state is eigenseparable. An interesting consequence of the above theorem is the following simple observation.

\begin{rem}
$E_\sep(\rho)=0$ is possible only if the kernel of $\rho$ is separable.
\end{rem}

So any state with an entangled kernel cannot be optimally locally measured. This is somewhat interesting, as the kernel of $\rho$ plays no role at all in the standard theory of separability. Separable states with entangled kernels provide insights into the difference between entanglement and the resource measured by $E_\sep$. 

For instance, take a nontrivial unextendible product basis $\{\ket{\phi_j} : j=1,\ldots,m\}$ in any composite system of Hilbert space dimension $D$, \ie~the $\ket{\phi_j}$ are orthonormal product vectors such that their simultaneous orthogonal complement contains no product vectors and has dimension $D-m \geq 1$ \cite{UPB}. It is evident that any state supported on the orthogonal complement is entangled, and on the other hand it is known that the projector onto it is PPT for any bipartition of the parties. That means that if we consider any state of the form $\rho = \sum_j p_j \ketbra{\phi_j}$ with $p_j>0$, its eigenprojections are definitely PPT; in fact, those with positive eigenvalues are separable, but the kernel projection is PPT entangled. Thus, by the above reasoning, invoking again Lemma~\ref{lemma:optimal-meas}, $E_\ppt(\rho) = 0 < E_\sep(\rho)$.

In a similar way, we can also separate $E_\ppt$ and $E_\rct$. Consider any Werner state in dimension $d$ \cite{werner1989states}, excluding the maximally mixed state. Its eigenprojections are the projectors onto the symmetric and the antisymmetric subspaces, respectively. While the symmetric projector is separable, the antisymmetric one is not even PPT. Since the sum of PPT operators must itself be PPT,
the antisymmetric projector cannot be expressed as a sum of PPT operators. By Lemma~\ref{lemma:optimal-meas}, this implies that $E_\ppt$ is strictly positive for every Werner state $\rho$ different from the maximally mixed state. In contrast, for $d>2$ both the symmetric and the antisymmetric projectors satisfy RCT. Therefore, applying Lemma~\ref{lemma:optimal-meas} again, we conclude that $E_\rct(\rho)=0<E_\ppt(\rho)$.

\section{Examples}
\label{sec:examples}
After the preceding general development of the theory, here we further analyze the various entropy gaps for specific examples of states. In particular, we consider Werner states, $n$-partite $W$ states, and classical-quantum states, and also consider how the gaps in certain 4-partite entangled states vary when partitions are changed or when subsystems are discarded.

\subsubsection{Werner states}
\label{subsec:werner}
For Werner states~\cite{werner1989states}, we analytically evaluate the entropy gap for any local class. In particular, we find that for any Werner state $\rho$, $E_\lostar(\rho)=E_\ppt(\rho)$, while, as we already noted at the end of Sec.~\ref{sec:SEP}, $E_\rct(\rho)=0$.

Consider a two-qudit system $AB$. A state $\rho_{AB}$ is a \textit{Werner state} if it is invariant under all unitaries of the form $U\otimes U$~\cite{werner1989states}. Denote by $F_{AB} =\sum_{i,j}\ket{i}\bra{j}_A\otimes\ket{j}\bra{i}_B$ the flip operator that exchanges $A$~and~$B$. Denote by $\Pi_\pm = \frac{1}{2}(\mathds{1} \pm F)$ the symmetric/antisymmetric projectors, $w_\pm = \tr \Pi_\pm = \frac{d(d\pm 1)}{2}$ the trace of these projectors, and  by $\rho_\pm = \Pi_\pm / w_\pm$ the states obtained by normalizing them. Every Werner state can be written as  
\begin{equation}  
    \rho_\lambda = (1-\lambda) \rho_+ + \lambda \, \rho_- ,  
\end{equation}  
with $\lambda\in[0,1]$. For these states we have the following result determining the entropy gaps.

\begin{prop}[Werner states]
    \label{thm:werner-gaps}
    Let $\rho_\lam$ be a \mbox{$d \times d$} Werner state and $\chi$ any set of POVMs such that \mbox{$\chippt \supset \chi \supset \chilostar$}. Then the optimal entropy for any $M \in \chi$ is given by $M_0 = (\Pi_D, \Pi_X)$, with projectors $\Pi_D = \sum_{i} \ketbra{ii}$ and $\Pi_X = \sum_{i \neq j} \ketbra{ij}$ projecting onto the diagonal and off-diagonal subspaces in any local basis. Therefore the gap is given by
    \begin{equation}
        E_\chi(\rho_\lam) = S_{M_0}(\rho_\lam) - S(\rho_\lam).
    \end{equation}
    Further, with $x_\lambda = \tr \Pi_X \rho_\lambda = (d + 2 \lambda -1)/(d+1)$, these entropies evaluate to
    \begin{equation}
    \begin{split}
        S_{M_0}(\rho_\lambda) &= h(x_\lambda) + (1-x_\lambda) \log d + x_\lambda \log d(d-1) \\
        S(\rho_\lambda) &= h(\lambda) + (1-\lambda) \log w_+ + \lambda\, \log w_-.
    \end{split}
    \end{equation}
\end{prop}

\begin{proof}
    
Since $\rho_\pm$ are proportional to projectors with orthogonal support, it follows that  
\begin{equation}  
    S(\rho_\lambda) = h(\lambda) + (1-\lambda) \log w_+ + \lambda\, \log w_-,  
\end{equation}  
where $h(\lambda)$ denotes the binary entropy. Denoting by $\Pi$ the measurement in the computational basis $\{\ket{ij}\}_{i,j}$ and by $(p_{ij})_{i,j}$ the corresponding outcome probabilities, it turns out that $p_{ij}$ depends only on whether $i=j$ or not. Consequently, the observational entropy obtained by measuring $\Pi$ takes the same value as measuring $M_0 = (\Pi_D, \Pi_X)$. Direct calculation then gives
\begin{equation}
    S_{M_0}(\rho_\lambda) = h(x_\lambda) + (1-x_\lambda) \log d + x_\lambda \log d(d-1),
\end{equation}
where $ x_\lambda = (d + 2 \lambda -1)/(d+1)$. Since $\Pi \in \chilostar$, and $S_{M_0}(\rho_\lam) = S_{\Pi}(\rho_\lam)$, we have that the gap is bounded by $E_\lostar(\rho_\lam) \leq S_{M_0}(\rho_\lam) - S(\rho_\lam)$. Note that by $U \otimes U$ invariance, the choice of the computational basis or another basis in defining $M_0$ is irrelevant. It remains to show that also $E_\ppt(\rho_\lam) \geq S_{M_0}(\rho_\lam) - S(\rho_\lam)$, which will complete the proof since $E_\lostar(\rho) \geq E_\ppt(\rho)$ in general.

Consider the ``twirl'' operation $T(\bullet)$, defined by the Haar integral $T(\bullet) = \int dU \; (U \otimes U) \bullet (U^\dag \otimes U^\dag)$
~\cite{werner1989states}. Applying the twirl to the POVM $M_0$ introduced above yields $\Pi'_D = T(\Pi_D) = d \rho_0$ and $\Pi'_X = T(\Pi_X) = d(d-1) \rho_{1/2}$.  
It can be verified that $M'=(\Pi'_D, \Pi'_X)$ constitutes another POVM that produces the same entropy as $M$ when applied to Werner states, due to the invariance under $U \otimes U$. Now, consider an arbitrary measurement $N$. The twirled measurement $N' = T(N)$ produces the same entropy when applied to Werner states. Since PPT-ness is preserved under local unitaries, if $N$ is PPT then $N'$ is also PPT. Each POVM element of $N'$ must therefore be proportional to a PPT Werner state. By~\cite{Chru_ci_ski_2006}, a Werner state is PPT if and only if it is separable. Hence, any POVM element of $N'$ must be proportional to a separable Werner state. Moreover, every separable Werner state $\rho_{\lambda}$ must satisfy $\lambda \leq 1/2$~\cite{werner1989states}, which implies that $\rho_{\lambda}$ can always be expressed as a convex combination of $\rho_0$ and $\rho_{1/2}$. Consequently, each element $N'_i$ can be written as a positive linear combination, $N'_i = c_{i|D} \, \Pi'_D + c_{i|X} \, \Pi'_X$. In order to satisfy $\sum_i N'_i = \mathds{1}$, it is necessary that $\sum_i c_{i|D} = \sum_i c_{i|X} = 1$. This implies that $N'$ is a classical post-processing of $M'$ in the sense of Def.~\ref{def:cpp} above, meaning there exists a stochastic map $\Lambda$ such that $N' = \Lambda M'$. By applying Thm.~\ref{thm:cpp-qpp-monotonicity}, we conclude that for any $N \in \chippt$,  
\begin{equation}  
    S_N(\rho_\lambda) = S_{N'}(\rho_\lambda) = S_{\Lambda M'}(\rho_\lambda) \geq S_{M'}(\rho_\lambda) = S_{M_0}(\rho_\lambda).
\end{equation}  
This shows that for any $N \in \chippt$, $S_N(\rho_\lam) \geq S_{M_0}(\rho_\lam)$. So $M_0$ is optimal, and $E_\ppt(\rho_\lam) \geq S_{M_0}(\rho_\lam) - S(\rho_\lam)$. 

Since $\chippt \supset \chi \supset \chilostar$, one has $E_\ppt \leq E_\chi \leq E_\lostar$. But since $E_\ppt(\rho_\lam) \leq E_\lostar(\rho_\lam)$ have the same value as upper and lower bounds, these values are all equal, and given by the value above. This completes the proof.
\end{proof}

Fig.~\ref{fig:werner-plots} shows plots of $E_\chi(\rho_\lambda)$ versus $\lambda$ for different Hilbert space dimensions, as well as a direct comparison of $S_{M_0}(\rho_\lam)$ and $S(\rho_\lam)$. One can observe that, coherently with Lemma~\ref{thm:convexity}, the gap $E_{M_0}(\rho)$ for fixed ${M_0}$ is convex over the set of states. The maximum gap is always attained at $\rho_1$, with a value of $E_\chi(\rho_1) = 1$ bit. This is consistent with the fact that $\rho_1$ is the furthest on the ``entangled side.'' The other extremal point corresponds to $\rho_0$, which satisfies $E_\chi(\rho_0) = 1 - 2/(d+1)$ bits. As discussed in Sec.~\ref{sec:SEP}, this provides an example of a separable state for which $E_\sep > 0$. One can also observe that the antisymmetric projector, an eigenprojector of the Werner states, is entangled, and therefore no Werner state can be eigenseparable (except in case that the symmetric and antisymmetric eigenvalues become degenerate).

\subsubsection{Genuineness and robustness in 4-partite entangled states}
As discussed in Sec.~\ref{sec:multipartite-picture}, a useful feature of the local $E_\chi$ gaps is that they can be compared meaningfully across different partitions. 

Here we leverage this possibility as a tool to analyze the genuineness and robustness of quantum correlations in a multipartite network. Genuineness~\cite{entanglement2009guhne} refers to the situation where correlation is ``genuinely'' shared among all parties of the network, and can be analyzed by studying the amount of correlation as a function of the fineness of partitioning. Robustness~\cite{dur2000three} refers to persistence of correlations in the presence of noise, or in particular to the loss of one of the subsystems, and can be analyzed by looking at the correlations in reduced density matrices of the system.

As an example we analyze these phenomena for several different multipartite qubit entangled states in a 4-partite system $ABCD$. The states we consider are a GHZ state~\cite{greenberger2007beyond}
\begin{equation}
	\ket{\phi_{\textrm{ghz}}} = \frac{\ket{0000}+\ket{1111}}{\sqrt{2}},
\end{equation}
a $W$ state~\cite{dur2000three}
\begin{equation}
	\ket{\phi_W} = \frac{\ket{1000}+\ket{0100}+\ket{0010}+\ket{0001}}{2},
\end{equation}
and a state containing two Bell pairs,
\begin{equation}
	\ket{\phi_{2\textrm{Bell}}} = \ket{\phi^+}_{AC} \otimes \ket{\phi^+}_{BD},
\end{equation}
where $\ket{\phi^+} = (\ket{00}+\ket{11})/\sqrt{2}$ denotes a single Bell pair~\cite{horodecki2009entanglement}. Each of these states has a different multipartite structure to its correlations, reflecting some of the diverse possibilities for multipartite entanglement/correlation~\cite{enriquez2016maximally}.\\

\begin{figure*}[t]
    \centering
    \begin{tabular}{cc}
        {\normalsize (a) Genuineness} & {\normalsize (b) Robustness}
        \\
         \includegraphics[width=\columnwidth]{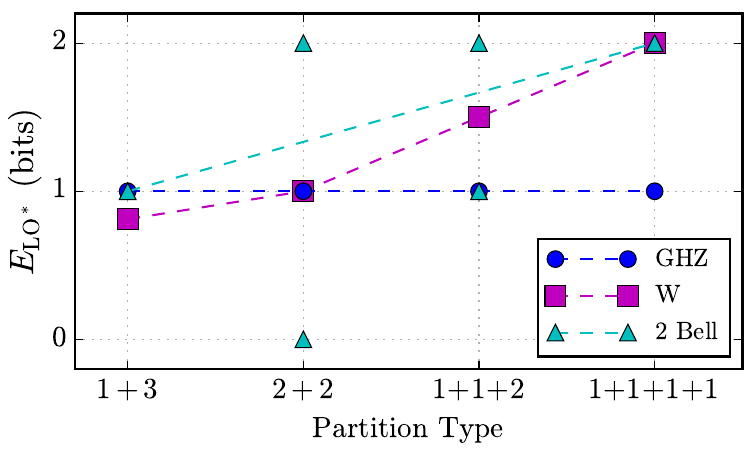}
         &
         \includegraphics[width=\columnwidth]{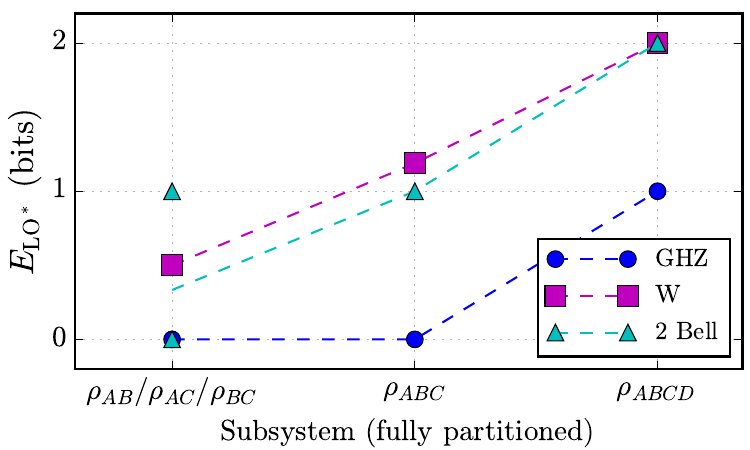}
    \end{tabular}
    
    \caption{Genuineness and robustness of quantum correlations in 4-partite GHZ, 2-Bell, and $W$ states. Plots show $E_\lostar$ in different partitions and when different subsystems are lost. Partitions are denoted by the type of splitting; \eg~``2+2'' partitions include $AB|CD$, $AC|BD$, and so on, while ``1+1+1+1'' means $A|B|C|D$. Subsystem loss is denoted by the reduced density matrix for which $E_\lostar$ is calculated; the remaining subsystems are fully partitioned as in the $1+ \ldots +1$ case. Markers show values for each particular choice of partition/reduction, while dashed lines show the average over all partitions/reductions of a given type. (a) Correlations in the $W$ state are more genuinely multipartite, as it takes fully partitioning to reveal the full amount of correlations. (b) Correlations in the $W$ state are most robust, decreasing least on average when subsystems are lost.}
    \label{fig:multipartite-genuine-robust}
\end{figure*}

In Fig.~\ref{fig:multipartite-genuine-robust}(a) we compute the $E_\lostar$ gap for each state across all possible partitions, while in Fig.~\ref{fig:multipartite-genuine-robust}(b) we evaluate the fully partitioned $E_\lostar$ gap for every reduced density matrix of the system.

For the GHZ state, the 1+1+1+1, 2+2 and 1+3 partitions follow directly from Thm.~\ref{thm:bipartite-pure-states}. For the 2+1+1 partition, we use Prop.~\ref{prop:part-mon} to lower bound $E_\lostar$ by that of the 2+2 partition, with the bound being saturated by a measurement in the computational basis, which is compatible with all partitions. For the two Bell pairs, the 1+1+1+1, 2+2, and 1+3 partitions are determined directly via Thm.~\ref{thm:bipartite-pure-states}, or by combining it with Eq.~\eqref{additivity} when uncorrelated subsystems are separated. Note that in the 2+2 partition, different choices of the same partition type can yield distinct values of $E_\lostar$. These individual values are represented by markers in Fig.~\ref{fig:multipartite-genuine-robust}(a), while the dashed lines indicate the average over all partitions of a given type. For the 1+1+2 partition, a lower bound from the 2+2 partition is again saturated by computational basis measurements. For the W states, the 1+3 and 2+2 partitions are computed using the corresponding Schmidt bases and Thm.~\ref{thm:bipartite-pure-states}, while $E_\lostar$ for the 1+1+1+1 and 2+1+1 partitions is obtained by numerically optimizing over local unitaries.

Regarding subsystem loss, losing one or two subsystems from the GHZ state results in a classically correlated state (Def.~\ref{def:classically-correlated-states}), implying $E_\lostar=0$ by Thm.~\ref{thm:LO-LOstar-zero-equivalence}. For the two Bell pairs, losing one subsystem, \eg, $D$, leaves $AC$ maximally entangled and uncorrelated from $B$, yielding $E_\lostar=1$ by Eq.~\eqref{additivity} and Thm.~\ref{thm:bipartite-pure-states}. This holds independently from the individual subsystem lost. Losing two subsystems can yield either $E_\lostar=1$ or $E_\lostar=0$: removing $BD$ (or equivalently $AC$) preserves correlations within $AC$ (or $BD$), giving $E_\lostar=1$, whereas losing initially uncorrelated subsystems (\eg~$A$ and $B$) leaves an uncorrelated state with $E_\lostar=0$. For W states, all $E_\lostar$ values under subsystem loss are once again determined numerically.

The plot reveals that the $W$ state appears to be both the most genuine and the most robust in terms of its 4-partite entanglement when compared with the other two states. Although correlations in the 2-Bell state are, on average, greater than or equal to those of the $W$ state (Fig.~\ref{fig:multipartite-genuine-robust}(a)), the latter's correlations vary more as additional partitions are added, revealing a greater genuineness of the correlations. On the other hand, the greater robustness is evident from Fig.~\ref{fig:multipartite-genuine-robust}(b), where we see that the $W$ state's correlations decrease least as subsystems are lost, on average over partitions/reductions.
Meanwhile, for the GHZ state, correlations are completely partition independent, and completely destroyed if any subsystem is lost. The \mbox{2-Bell} state has the most complex structure: the amount of correlations depends strongly on the specific partition, showing that while all four parties share correlations, they do so asymmetrically. This is why the 2+2 partition in Fig.~\ref{fig:multipartite-genuine-robust}(a) yields an average $E_\lostar$ greater than 1. If we consider the bipartition where the two parts are not entangled, that is $AC|BD$, we get $E_\lostar=0$; for the other two possible bipartitions, however, we obtain a maximally entangled state between two two-qubit systems, which gives $E_\lostar=2$.

This raises another aspect of genuineness, quantified by the amount of permutation invariance in correlations, describing how symmetrically all the parties participate. A detailed exploration between genuineness and symmetry of multipartite states is left for future work.

\subsubsection{Classical-quantum states}
Classical-quantum (CQ) states, defined as those of the form (see \eqref{eqn:CQ-state})
\begin{equation}
\label{eqn:CQ-state-2}
    \rho = \sum_k \lambda_k \, \ketbra{k} \otimes \rho_k,
\end{equation}
provide an easy set of states for which the LO*, LO, and LOCC entropies can be evaluated. A fundamental property of CQ states, that the optimal LO/LO* measurements must measure in the classical basis on the C side, was proved already in Thm.~\ref{thm:CQ-state_LO-LOstar}. Moreover, it was shown in \eqref{eqn:CQ-ensemble-mutual-info} how these LO/LO* gaps relate to accessible information~\cite{wilde2011notes} in the corresponding ensemble. Here we continue the study of CQ states in somewhat more detail. 

Consider a CQ state \eqref{eqn:CQ-state-2} on bipartite system $AB$. A first observation is the reduction of the problem to an optimization in the quantum side (system~$B$).

\begin{prop}
\label{prop:cl-quantum}
    For CQ state $\rho = \sum_k \lam_k \ketbra{k} \otimes \rho^B_k$, for either $\chi=\chilostar$ or $\chi=\chilo$,
    \begin{equation}
        E_{\chi}(\rho) = \inf_N \,  \sum_k p_k \big(S_N(\rho^B_k) - S(\rho^B_k)\big),
    \end{equation}
    where for LO* the infimum is over all projective measurements $N$ on system $B$, and for LO the infimum is over all POVMs $N$ on system $B$.
\end{prop}

\begin{proof}
By Prop.~\ref{thm:CQ-state_LO-LOstar}, for LO/LO* $S_\chi(\rho) = \inf_N S_{\Pi_\cb \otimes N}(\rho)$ with $N$ infimized over the appropriate set. Evaluating this gives $S_\chi(\rho) = H(\lambda_k) + \inf_N  \sum_k \lam_k S_N(\rho^B_k)$, using the chain rule \eqref{eqn:qpp-chain-rule}, where $H(\lam_k)$ is Shannon entropy of the coefficients. Meanwhile, using~\cite[Thm.~11.2.2]{wilde2011notes} one finds $S(\rho) = H(\lam_k) + \sum_k \lam_k S(\rho^B_k)$. Therefore $E_\chi(\rho) = S_\chi(\rho) -S(\rho)$ is of the claimed form.
\end{proof}

This provides a simple way to optimize LO/LO* measurements on CQ states. For LOCC, the result is even simpler. Indeed, CQ states are not correlated enough to have an LOCC or separable gap.

\begin{prop}
    If $\rho$ is CQ then $E_\locc(\rho)=0$.
\end{prop}
\begin{proof}
    One can easily obtain an optimal LOCC measurement. First, the classical system is measured in the $\ketbra{k}$ basis. Then, for whichever $k$ is obtained, the quantum system is measured in the $\rho_k$ eigenbasis. Using the chain rule \eqref{eqn:qpp-chain-rule}, this gives $S_M(\rho) = H(\lam_k) + \sum_k \lam_k S(\rho_k)$. But comparing with the proof of Prop.~\ref{prop:cl-quantum}, this is the same value as $S(\rho)$, so $E_\locc(\rho) = 0$.
\end{proof}

We can now come back to the example of the 2-qubit state $\rho = \frac{1}{2}\left(\ketbra{00} + \ketbra{1+}\right)$ that was earlier claimed to have an LO* gap of $E_\lostar(\rho) = 0.5$ bits. Previously, in \cite{schindler2020correlation}, this was calculated numerically, with an analytical upper bound given by $\alpha \approx 0.601$ bits. With the help of Prop.~\ref{prop:cl-quantum} above, we can now evaluate the value of $0.5$ bits analytically. The conditional states in Prop.~\ref{prop:cl-quantum} become $\rho_0^B = \ketbra{0}$ and $\rho_1^B = \ketbra{+}$, which are both pure. Thus $E_\lostar(\rho) = \frac{1}{2}\big(S_N(\ketbra{0}) + S_N(\ketbra{+})\big)$, optimized over the LO* measurements $N$. Considering $N$ as the measurement along the axis $\hat{n} \equiv (n_x, n_y, n_z)$. We find that both $S_N(\ketbra{0})$ and $S_N(\ketbra{+})$ are minimized when $\hat{n}$ lies in the $xz$-plane. Restricting our analysis to such versors, $S_N(\ketbra{0}) + S_N(\ketbra{+})$ becomes a function of $n_z$ alone. This function attains its minimum for $n_z = 0$, or $n_z\pm1$, yielding $E_\lostar(\rho) = 0.5$. Notably, it is impossible to minimize both conditional entropies simultaneously. Indeed, $S_N(\ketbra{0})$ is minimized for $n_z=\pm1$, while $S_N(\ketbra{+})$ for $n_z=0$. Similarly, neither the marginal nor mutual terms in~\eqref{eqn:marginal-mutual} can be minimized at the same time. This illustrates a case where $E_\lostar$ differs from other related measures discussed in~\cite[VII]{schindler2020correlation}.

Instead of $\rho = \frac{1}{2}\left(\ketbra{00} + \ketbra{1+}\right)$, one could also consider the pure state $\ket{\psi}=\frac{1}{\sqrt{2}}\left(\ket{00}+\ket{1+}\right)$ constructed by superposing instead of mixing. For $\psi = \ketbra{\psi}$ one finds, with Cor.~\ref{thm:bipartite-pure-states}, that $E_{\lostar}(\psi) \approx 0.601$ bits, a strictly larger value than for the related CQ state. One can observe that $\rho$ is obtained from this pure state by the dephasing channel $\Phi = \Phi_\cb \otimes \one$, where $\Phi_\cb(\bullet) = \sum_k \ketbra{k} \bullet \ketbra{k}$ is a dephasing in the classical basis. The fact that such a map decreases the LO/LO* entropy can be generalized to CQ states of arbitrary two qudit systems.

\begin{prop}
    Consider bipartite pure state $\psi = \ketbra{\psi}$ on system $AB$ with state vector
    \begin{equation}
        \ket{\psi}=\sum_k c_k \, \ket{k} \otimes \ket{\psi_k},
    \end{equation}
    where $\ket{k}$ are an orthonormal basis in system $A$ and $\ket{\psi_k}$ are arbitrary vectors in system $B$. Let $\Phi = \Phi_\cb \otimes \one$, with  $\Phi_\cb(\bullet) = \sum_k \ketbra{k} \bullet \ketbra{k}$, denote the locally dephasing channel in the $\ket{k}$ basis of system $A$. The state $\rho = \Phi(\psi)$ arising from this channel is the CQ state
    \begin{equation}
        \rho=\sum_k \lam_k \, \ketbra{k} \otimes \psi_k,
    \end{equation}
    where $\psi_k = \ketbra{\psi_k}$ and $\lam_k = |c_k|^2$.
    Let $\chi \in \{\text{LO},\text{LO*}\}$, then
    \begin{equation}
        E_\chi(\rho)\leq E_\chi(\psi).
    \end{equation}
\end{prop}

\begin{proof}
    Both $\rho$ and $\psi$ have the same reduced density matrices $\rho_B = \psi_B = \sum_k \lam_k \psi_k$ in system $B$, and it holds by Cor.~\ref{thm:bipartite-pure-states} that $E_{\chi}(\psi) = S(\psi_B)$ for both LO and LO*. Meanwhile, for $E_\chi(\rho)$, from Prop.~\ref{prop:cl-quantum} we have that
    \begin{equation}
        \begin{split}
            E_\chi(\rho) 
            &=\inf_N \textstyle\sum_k \lam_k (S_N(\psi_k)-S(\psi_k) )
            \\
            &= \inf_N \textstyle\sum_k \lam_k S_N(\psi_k)
            \\
            &\leq \inf_N \textstyle S_N(\psi_B)
            \\
            &= S(\psi_B).
        \end{split}
    \end{equation}
    where we have first used the fact that the $\psi_k$ are pure, then concavity of the entropy obtaining the inequality, then Lemma~\ref{lemma:optimal-meas} stating that the infimum (over POVMs or projective measurements) is the von Neumann entropy. But it was already shown that $S(\psi_B) = E_\chi(\psi)$, and therefore we have $E_\chi(\rho) \leq E_\chi(\psi)$, completing the proof.
\end{proof}

This provides additional perspective on the question of $E_\chi$ monotonicity raised in Secs.~\ref{sec:monotonicity-general}~and~\ref{sec:entanglement-discord-and-re-measures}. The map $\Phi = \Phi_\cb \otimes \one$ is unital and local, and is therefore compatible with the LO class. This is the class of channels where monotonicity for $S_\lo$ is known to hold, as in Thms.~\ref{thm:channel-monotonicity-general}~and~\ref{thm:channel-monotonicity-local-separable}. This example shows a class of cases where $E_\lo$ is also monotone under such maps.

\subsubsection{W states}
The $3$-partite $W$ state is defined as $W_3=\ketbra{W_3}$, with $\ket{W_3} = \sqrt{1/3} \big(\ket{100} + \ket{010} + \ket{001})$, and is widely studied in entanglement theory as an entangled state not LOCC interconvertible with the GHZ state~\cite{dur2000three}. In contrast to the GHZ state, evaluating the entropy gaps for $W_3$ is considerably more challenging: so far we have derived an analytic expression only for the PPT class. {For $E_\lostar$ and $E_\lo$ we have obtained numerical values; however, an analytic proof of these results remains elusive. As for the LOCC and SEP classes, we currently have only upper and lower bounds, leaving the exact gap undetermined. Resolving this gap remains an intriguing direction for future research.}

We approached the numerical computation of $E_\lostar$ by randomly sampling local bases according to the Haar measure on each qubit, and then calculating the OE resulting from measurements in the tensor product of these bases. In particular, we extended this analysis to the $n$-partite $W$ state $W_n=\ketbra{W_n}$, with $\ket{W_n} = \sqrt{1/n}(\ket{10 \ldots 0} + \ldots + \ket{00 \ldots 1})$ being a uniform superposition of all computational basis states with exactly one qubit in state $1$ and the others in state $0$. Numerical results suggest that, for all $n$,
\begin{equation}
\label{eq: elostar_Wn}
     E_\lostar(W_n) = \log n,
\end{equation}
with the optimal measurement being in the computational basis. This is consistent with the result in~\cite{modi2010unified}, which reports the value $\log 3$ for $W_3$. Moreover, during our numerical exploration, we observed a more general pattern that implies Eq.~\eqref{eq: elostar_Wn} as a special case: any linear combination of computational basis states {containing the same number $k$ of $1$’s  
appears to have minimal OE when measured in the computational basis.
Indeed, for 
\[
  \ket{\psi} = \sum_{\substack{i_1+i_2+i_3=k \\ i_j\in\{0,1\}}} \alpha_{i_1i_2i_3} |i_1i_2i_3\rangle
\] 
numerical evidence suggests that 
\[
  E_\lostar(\psi) = -\sum_{\substack{i_1+i_2+i_3=k \\ i_j\in\{0,1\}}} |\alpha_{i_1i_2i_3}|^2 \log |\alpha_{i_1i_2i_3}|^2.
\]
These findings suggest that a central role is played by the invariance under conjugation with unitaries of the form $V = (e^{i\theta \sigma_z})^{\otimes n}$, and a possible extension to Dicke states~\cite{dicke1954coherence}.}

Regarding $E_\lo$, we exploit the fact that OE is concave in the POVM. This allows us to restrict our attention to extremal POVMs, \ie, those that cannot be expressed as a convex combination of other POVMs. It is well-known that extremal POVMs for a single qubit contain at most four elements, and -- except for the trivial POVM consisting only of the identity -- these elements must be rank-one projectors. This result significantly simplifies the numerical implementation of $E_\lo(W_3)$.
We find that no LO measurement yields an OE below $\log 3$, indicating that there is no separation between $E_\lo$ and $E_\lostar$ for the state $W_3$. In contrast, a separation does arise when we consider the LOCC class. Specifically, by performing the projective measurement $\{\alpha\ket{0} + \beta\ket{1}, \beta^*\ket{0} - \alpha^*\ket{1}\}$ on the first qubit, with $|\alpha|^2 = 1/3$, followed by a conditional measurement on the remaining two qubits—designed to achieve the entanglement entropy of the resulting post-measurement state—we obtain a {minimal OE of $\approx 1.550 < \log 3 \approx 1.585$. 
This value serves as an upper bound for $E_\locc$ and, by Eq.~\eqref{eqn:chain-Echi}, also for $E_\sep$. A lower bound is provided by $E_\ppt$, which we are able to compute analytically.}

Let $M=(M_i)_i$ be a PPT POVM (see Sec.~\ref{sec:local-measurement-classes}). The state $W_3$ is invariant under conjugation with local phase unitaries {$V=(e^{i\theta\sigma_z})^{\otimes 3}$} and permutation matrices $U^{(\pi)}$ implementing permutations $\pi$ of three objects. As a consequence, the same probability distribution $(p_i)_i$ is obtained whether we measure $M$ or $VMV^{\dagger}$. Since convex combinations of these rotated POVMs also yield the same probability distribution, we can perform a twirl over the unitaries $V$ belonging to the group generated by permutations and local phase unitaries, and always obtain a POVM invariant under conjugation with such $V$ that produces the same OE. We are thus led to characterize the set of operators invariant under conjugation with $V$, and among these, identify those that satisfy the PPT condition. In particular, the invariant operators can be shown to be linear combinations of six projectors:
$Q_1=\ketbra{W_3}$, $Q_2=\ketbra{\overline{W}_3}$, $Q_3=P_1-W_3$, $Q_4=P_2-\overline{W}_3$, $Q_5=\ketbra{000}$, and $Q_6=\ketbra{111}$, 
where $\ket{\overline{W}_3}$ is the anti-W state defined as $\ket{\overline{W}_3}\equiv\sqrt{1/3}(\ket{011}+\ket{101}+\ket{110})$,
$P_1=\ketbra{100}+\ketbra{010}+\ketbra{001}$ and $P_2=\ketbra{011}+\ketbra{101}+\ketbra{110}$.
Note that an outcome has nonzero probability only if the corresponding POVM element has a nonzero coefficient for $Q_1$. This allows us to restrict attention to POVMs $M = (M_i)_i$ whose elements take the form
\begin{equation}
    M_i = p_i Q_1 + \sum_{a=2}^6 q^{(i)}_a Q_a,
\end{equation}
where $\sum_a q^{(i)}_a = 1$, and $p_i$ are positive numbers such that $\sum_i p_i = 1$. We have $p_i = \tr\big(M_i W_3\big)$, and therefore
\begin{align}
    S_M(W_3) &= -\sum_i p_i \log \frac{p_i}{\tr M_i} \\
    &= \sum_i p_i \log\left(1 + \sum_a \frac{q^{(i)}_a}{p_i} \tr Q_a \right).
\end{align}
The expression inside the logarithm is the trace of the operator $M' = Q_1 + \sum_a t_a Q_a$, where $t_a\equiv q^{(i)}_a / p_i$. This leads to a lower bound for $E_\ppt(W_3)$:
\begin{equation}
\label{eq:low_bound_ppt}
    E_\ppt(W_3) \geq \log\left( \min_{M' \in \chi} \tr M' \right),
\end{equation}
where $\chi$ is the set of operators $M' = Q_1 + \sum_a t_a Q_a$ that are positive and PPT with respect to the first qubit (due to permutation symmetry, it is sufficient to consider the partial transpose only on the first qubit). In particular, if we can find an operator $M' \in \chi$ such that $t_a \leq 1$ for all $a$, and such that $\one - M'$ is also PPT, then the lower bound in Eq.~\eqref{eq:low_bound_ppt} is tight, and the POVM $(M', \one - M')$ is optimal for the PPT class. We have calculated that the operator $M'$ in $\chi$ with the minimal trace corresponds to the following coefficients:
\begin{equation}
    t_2 = \frac{1}{2}, \ t_3 = t_4 = 0, \  t_5 = \frac{2}{3}, \  t_6 = \frac{1}{12},
\end{equation}
which yields a trace of $9/4$. Since all coefficients are less than 1 and $\one - M'$ is PPT, the lower bound in Eq.~\eqref{eq:low_bound_ppt} is in fact achievable within the PPT class, and the POVM $(M', \one - M')$ is optimal.
In conclusion, we find that $E_\ppt(W_3) = \log \frac{9}{4} < \log 3$. Proving that $(M', \one - M')$ belongs to the SEP class—\ie, showing that both $M'$ and $\one - M'$ are fully separable—would yield an analytic value for $E_\sep(W_3)$ as well. However, this does not appear to be a straightforward task. We attempted to determine whether $M'$ is entangled by employing higher levels of the DPS hierarchy~\cite{Doherty_2004} and by searching for explicit entanglement witnesses~\cite{horodecki1996separability}. Nevertheless, none of the tests conducted so far have succeeded in detecting entanglement, leaving open the question of whether $(M', \one - M')$ belongs to the SEP class. 

As previously mentioned, $E_\ppt(W_3)$ still provides a valid lower bound for both $E_\locc(W_3)$ and $E_\sep(W_3)$. Although a definitive analytical or numerical determination of these two quantities remains open, our results establish that their values must lie within the interval between $\log \frac{9}{4} \approx 1.170$ and $\approx 1.550$. 

\section{Conclusions}
\label{sec:conclusions}
We have explored the hypothesis that the observational entropy gap $E_\chi(\rho)$ of a multipartite state $\rho$ contains information about its multipartite entanglement, for classes $\chi$ of POVMS tied to the locality of the composite system, such as LO*, LO, LOCC, SEP, and PPT.

We show that these gaps are not entanglement monotones (local operations or LOCC can increase them from zero to a positive quantity), but do provide a lens into the quantum correlations in a system. In particular the gaps are related to the limitations of local observers to extract information through measurements, and we find that these gaps are bounded below by the relative entropy of entanglement, and above by the relative entropy of quantumness (which is equal to the LO* gap). More encouragingly even, for pure bipartite states, all of the entropy gaps above equal the entropy of entanglement, the distinguished entanglement measure in this setting.

It remains an open question whether the entropy gaps $E_\chi$ are monotones in \emph{some} resource theory, possibly a restriction of LOCC. As we have found that in general, $E_\lostar < E_\lo < E_\locc < E_\sep < E_\ppt$, already for bipartite mixed states, and plausibly for tripartite pure states, the next question would be which resource aspect each of them quantifies. Whatever the resource picture, it appears to be tied to the interplay between purity and entanglement.

Our findings open up several other intriguing avenues for further exploration.
Regarding separable measurements, a refinement of the Separable Measurement Theorem \ref{thm:sep-bound}) would be highly desirable -- in particular one that establishes a lower bound in terms of separable measurements on subsystems $A$ and $B$, if these are themselves composite systems.

Given the entropy gaps' demonstrated effectiveness in capturing the genuineness and robustness of multipartite entangled states, it becomes interesting to quantify the degree of permutation invariance in the correlations. This would illuminate the extent to which all parties participate symmetrically.

Due to the apparent simplicity of $W$ states, it is compelling to precisely characterize the entropy gap across diverse measurement classes and for arbitrary number of parties. Notably, $W$ states are also candidate pure states that may distinguish between LO and LOCC operations. The broader question of whether multipartite pure states can separate the various operational classes remains open.

Finally, although a few results are also applicable to infinite-dimensional systems, the key properties of the entropy gap in such systems remain for future study.


\begin{acknowledgments}
The authors are grateful to John Kimble for pedagogical advice on the effective presentation of OE, although in the end we didn't adopt his ``ferret'' approach.
LR is grateful to the Department de Fisica, Universitat Aut\`onoma de Barcelona, for the kind hospitality during the first period of this work.
SM acknowledges financial support from ``PNRR MUR project PE0000023-NQSTI''.
AW and JS are supported by the Spanish MICIN (project PID2022-141283NB-I00) with the support of FEDER funds, by the Spanish MICIN with funding from European Union NextGenerationEU (PRTR-C17.I1) and the Generalitat de Catalunya, and by the Spanish MTDFP through the QUANTUM ENIA project: Quantum Spain, funded by the European Union NextGenerationEU within the framework of the ``Digital Spain 2026 Agenda''. 
AW was or is furthermore supported by the European Commission QuantERA grant ExTRaQT (Spanish MICIN project PCI2022-132965), by the Alexander von Humboldt Foundation, and by the Institute for Advanced Study of the Technical University Munich.
JS acknowledges support by a Beatriu de Pin\'os postdoctoral fellowship \mbox{(2022 BP 00181)} of the Ministry of Research and Universities of the Government of Catalonia under EU Horizon 2020 MSCA grant agreement No 801370.
\end{acknowledgments}





\bibliography{biblio}


\end{document}